\documentclass{aa}
\usepackage{graphicx}

\usepackage{txfonts}
\usepackage{natbib}
\bibpunct{(}{)}{;}{a}{}{,}
\usepackage[figuresright]{rotating}
\newcommand{\beq}{\begin{equation}}
\newcommand{\eeq}{\end{equation}}

\newcommand{\myarcsec}{\hbox{$.\!\!^{\prime\prime}$}}
\newcommand{\myarcmin}{\hbox{$.\!\!^{\prime}$}}

\begin{document}
   \title{The Shapley super-cluster}

   \subtitle{New X-ray detections and mass distribution}

   \author{E. De Filippis
          \inst{1,2,3}
          \and
          S. Schindler\inst{4}
          \and
          T. Erben\inst{5}
          }

   \offprints{E. De Filippis, \email{betty@na.infn.it}}

   \institute{Dipartimento di Scienze Fisiche, Universit\`{a} degli 
		Studi di Napoli ``Federico II'',
                Via Cinthia 9, Compl. Univ. Monte S. Angelo,
                80126 Naples, Italy\\
		\email{betty@na.infn.it}
         \and
		MIT Kavli Center for Astrophysics and Space Research,
               Massachusetts Institute of Technology,
               70 Vassar Street, Building 37,
               Cambridge, MA 02139, USA
	\and
		Astrophysics Research Institute, 
		Liverpool John Moores University,
		Birkenhead CH41 1LD,
                United Kingdom
	\and
		Universit\"at Innsbruck,
		Institut fuer Astrophysik,
		Technikerstr. 25,
		A-6020 Innsbruck, Austria\\
		\email{Sabine.Schindler@uibk.ac.at}
	\and
		Institut f\"ur Astrophysik und Extraterrestrische 
                 Forschung (IAEF),
		 Universit\"at Bonn,
		 Auf dem H\"ugel 71,
		 D-53121 Bonn, Germany\\
		\email{terben@astro.uni-bonn.de}
             }

   \date{Received 22/06/05; accepted 05/08/2005}

   \abstract{
The largest and the deepest super-structure known today is the Shapley super-cluster.
This is the sky area with the highest over-density of galaxy clusters and
therefore also an ideal region to test the effects of a high density environment on 
galaxies and on clusters. \\
We performed an X-ray survey of a wide region surrounding the Shapley 
super-structure. 
Additionally to previously known super-cluster X-ray members,
we identified diffuse X-ray emission from 
$35$ cluster candidates without previous X-ray detection. $21$ of them were
previously known, optically selected super-cluster members, while the
other candidates had not been previously detected in any wavelength
range. Optical follow-up observations revealed that at least four of 
these new
candidates also have optical cluster counterparts.
The super-cluster shows a slightly flattened and elongated morphology. 
Clusters outside the central dense core are preferentially located in four 
perpendicular filaments in a similar way to what is seen in simulations of Large
Scale Structure.\\
We measure the cluster number density in the region to be more than
one order of magnitude higher than the mean density of rich Abell
clusters previously observed at similar Galactic latitudes; this
over-density, in the super-cluster outskirts, is mainly due to an excess 
of low X-ray luminous
clusters (with respect to an average
population), which leads us to think that the whole region is
still accreting low luminosity, small objects 
from the outskirts.
Pushing our total X-ray mass estimate to fainter clusters 
would drastically increase the total super-cluster mass measure, 
because of the presence of the rich X-ray low luminosity population.

   }

   \maketitle
%

\section{Introduction}
Galaxy clusters are commonly identified as the largest virialized
structures in the Universe. Clusters are themselves embedded in larger
systems, extending to tens of Mpc; the cores of a few of these
super-structures have exceptionally high cluster number densities, and
hence are argued to be collapsing under the effect of
their own gravity. Detailed measurements of the size, morphology and
mass of these collapsing regions are consequently of profound
cosmological importance and for a correct understanding of the
large-scale structure in the Universe. A considerable observational
effort has been devoted in the past twenty years to the understanding
of one of the densest and richest aggregations of galaxy clusters: the
Shapley super-cluster (SC). This effort has provided a deep knowledge
of the internal structure and of the dynamics of the central and
brightest clusters and aggregations of clusters in the region.

However, a complete overview and analysis of the whole area in the X-rays 
is still missing; because of the large angular size only
specific regions have been analyzed so far. Our new analysis
suggests that previous studies did not include a large portion of the SC 
clusters because of their high flux limit.
Therefore previous determinations of the matter
over-density in this region were certainly underestimations and hence
also the cosmological conclusions drawn from these analyses suffered from
bias.

The aim of this work is to give a more complete overview of the X-ray
properties of extended sources in this exceptionally rich and crowded
area of the sky.  The paper is organized as follows.  Basic
information on the Shapley super-cluster are given in
\S~\ref{sec:Shapley}. The surveyed sky area is described 
in \S~\ref{sec:RASS3}.
In \S~\ref{sec:Algorithm} we give details of
our new algorithm, written to detect extended
structures without any a priori model assumption. 
In \S~\ref{sec:cl_detection} details are given on additional selection
criteria applied to our algorithm in order to discriminate non-cluster
sources. 
Lists and tables with the resulting detected clusters and
their properties can be found in \S~\ref{sec:fin_res}, together with
results on the efficiency of our detection algorithm and a
case by case discussion for all clusters
undetected in our survey that had previous X-ray
detections. In this section we also describe our second step analysis in which we
perform a deeper search, to fainter limits, for optically known
clusters with no X-ray detection. 
In \S~\ref{sec:new_det} the new cluster candidates detected in our survey
and their optical follow-up observations are described. 
The cluster distribution and the cluster number density in the region, 
results from the analysis of their X-ray luminosity function and of their 
cumulative mass profiles, together with optical versus 
X-ray cluster properties and a discussion on merger rate
are given in
\S~\ref{sec:morph_anal},~\ref{sec:XLF},~\ref{sec:mass}, respectively. 
\S~\ref{sec:results} gives a summary and
discussion of the results.

Throughout this paper we quote errors at the $90\%$ confidence level and, unless otherwise stated, we use $H_0=72\ \mathrm{km\ s}^{-1} \mathrm{Mpc}^{-1}$ ($\Omega_m=0.3$, $\Omega_{\Lambda}=0.7$).


\section{The Shapley super-cluster}
\label{sec:Shapley}
By the end of the 1980s a considerable amount of evidence suggested
the existence of a small but systematic and significant perturbation
on the smooth Hubble flow. \citet{Bur86} observed some form of
streaming for galaxies within $60\ h^{-1}\ {\rm Mpc}$ of the Local
Group toward the Centaurus SC in the Southern
Hemisphere. It was originally considered a general streaming motion,
but then it was believed that these velocities were caused
by a {\it Great Attractor} in the direction of the Centaurus SC
when~\citet{Sca89} reported the presence of a very rich concentration
of clusters of galaxies, centered at RA$=13^{\rm h}05^{\rm m}57\fs
8$, Dec$=-33\degr04\arcmin03\farcs 0$ (J2000). This concentration
was estimated to have a distance ranging from $30\ h^{-1}\ {\rm Mpc}$
to $200\ h^{-1}\ {\rm Mpc}$ with a central peaked component at $145\
h^{-1}\ {\rm Mpc}$. About $28$ clusters were determined to belong to
this concentration in about $2.5\times10^5\ h^{-3}\ {\rm Mpc}^3$,
producing a number over-density of clusters of more than a factor of
ten with respect to the mean density of Abell clusters at similar
galactic latitudes.\\ \citet{Ray91} reported
that the Shapley concentration was not only the most remarkable
feature that appears when looking at the spatial distribution of Abell
clusters, but that it was also the richest SC in the sky in terms of
X-ray emitting clusters.\\ The discovery of this large concentration
of clusters, together with the coherent deviations from the Hubble
flow in the direction of Hydra-Centaurus observed in nearby parts of
the Universe, led to the hypothesis that the two phenomena could in
some way be related.\\ In the following years many of the cluster
galaxies in the central area of the region were observed
spectroscopically~\citep{Bar94,Qui95,Qui97,Bar98,Bar00}. Several of
them were also observed in radio wavelengths~\citep{Nik95,Ven97,Rei98}
and many ROSAT pointed observations were devoted to clusters belonging
to this superstructure~\citep{Bar96,Sch96,Ett97,Dav99}. Some of these
objects were observed also with other X-ray satellites such as
ASCA~\citep{Mar97,Mar98a,Han99}, Beppo-SAX~\citep{Ett00,Bon01,Nev01},
Einstein~\citep{Dav93,Jon99} and Ginga~\citep{Day91}.
Chandra and XMM-Newton were used to observe some of the Shapley
clusters~\citep{Gas03}, confirming the strong presence of merging
events in the area.\\


\section{A complete X-ray analysis of the Shapley area}
\label{sec:RASS3}
Originating in the hot intra-cluster gas filling the cluster
potential, the X-ray emission provides the means to assess the size,
shape and mass of a cluster, offering at the same time a unique
opportunity to efficiently detect and to characterize galaxy clusters.

Our aim is to detect and analyze extended sources in a wide region
surrounding the Shapley SC. First, we want to detect all known
Shapley clusters already confirmed as X-ray emitters. We then aim to
verify if we can detect any diffuse X-ray emission from known Shapley
clusters that do not yet have any X-ray detection. Finally, we intend
to check for the possible presence of unknown extended sources in the
area. The application of a uniform detection technique allows a direct
comparison of the X-ray properties of all the above sources.

The only possible way to perform a coherent analysis of such a wide
area of the sky is to have the whole of it observed with the same
instrument and for a reasonably similar amount of time. This kind of
data is unfortunately not available from the newest X-ray
satellites. The ROSAT All-Sky Survey (RASS) therefore remains
a unique archive of data for this type of analysis.

We analyze a region centered on RA$=13^{\rm h}20^{\rm m}0\fs 0$,
Dec$=-33\degr00\arcmin00\farcs 0$ (J2000). The total solid angle
covered is $\Omega=0.27\ {\rm sterad}$; its size is chosen such that
all known cluster members are included. A large border around the SC
area is also included in the analysis to allow for a study not only of
its central regions, but also of its outskirts.

\begin{figure}[ht]
\centering
\resizebox{\hsize}{!}{\includegraphics{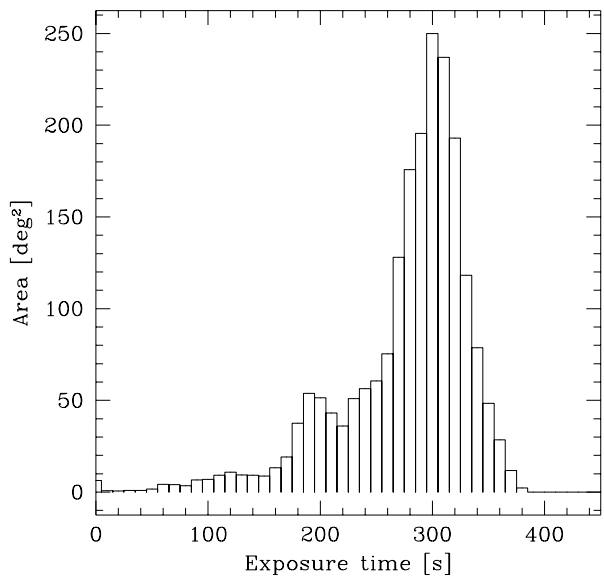}}
\caption{RASS-III exposure time distribution in the analyzed area.}
\label{fig:exp_plot}
\end{figure}

For this analysis we use the third re-processing of the RASS data,
released on March 22nd, 2000 (RASS-III; data can be down-loaded via
anonymous ftp from ftp.xray.mpe.mpg.de). $25$ plates are analyzed in
order to cover the whole selected area. Each field is a superposition
of several scans; the exposure time therefore varies from field to
field and within each field. A histogram of the exposure time
distribution over the whole area is shown in
Fig.~\ref{fig:exp_plot}. The mean and median values of the exposure
time are $277\ {\rm s}$ and $293\ {\rm s}$, respectively.

\subsection{Clusters in the surveyed area}
\label{sec:known_cl}
In the surveyed area there are $768$ optically known galaxy
clusters. Part of them are grouped to form two large superstructures:
the Hydra-Centaurus and the Shapley SCs.\\ Hydra-Centaurus is located
at RA$=13^{\rm h}13^{\rm m}35\fs 1$,
Dec$=-33\degr21\arcmin53\farcs0$ (J2000.0) at a redshift of
$z=0.014$. It was believed to be composed by five clusters: Abell
3526, Abell 3537, Abell 3560, Abell 3565 and Abell 3574~\citep{Ein94},
but Abell 3537 and Abell 3560 were later shown to be at higher
redshifts. More recently two other clusters were spectroscopically
identified as probable SC members: Abell S0753 and Abell 3581.\\ The
Shapley SC lies behind Hydra-Centaurus, almost in the same projected
area of the sky, with its center located at RA$=13^{\rm h}05^{\rm
m}57\fs 8$, Dec$=-33\degr04\arcmin03\farcs0$ (J2000.0) and
$z=0.043$.\\ 
$174$ of the $768$ known clusters in the area have no measured
redshift. The other $594$ are located between $z=0.0114$ (Centaurus
clusters) and $z=0.85$ (LCDCS 0831); only five of them are
closer than $z=0.03$, while $542$ are at $z>0.07$. The remaining $47$
clusters are in the intermediate redshift range $0.03 \leq z \leq
0.07$ and hence potential members of the Shapley SC. These
we analyze in more detail.
We start selecting such a wide redshift
range to be sure not to exclude any cluster belonging to the Shapley
SC and also (as we did for the sky area to be surveyed) to enable an
analysis of the neighboring regions.

The $47$ clusters which lie inside the surveyed area and within the selected redshift range are:
\begin{itemize}
\item Abell 3558, Abell S0740, Abell 3570, Abell 3575, Abell S0758,
Abell 3571, Abell 3578, SC 1336-314, Abell S0757, SC 1329-314, Abell
1736, Abell 3559, Abell 1631, Abell 3554, RX J1332.2-3303, Abell 1644,
Abell S0718, Abell 3556, SC 1342-302, Abell 3553, Abell 3555, Abell
3560, Abell S0721, Abell 3562, Abell S0724, Abell 3577, SC 1327-312,
Abell S0729, Abell 3544, Abell S0734, Abell S0731, Abell 3564, Abell
S0742, Abell 3566, Abell 3568, Abell 3572, Abell 3552, Abell 1709,
Abell 3542, Abell 3537, Abell 3528, RX J1252.5-3116, Abell 3530, Abell
3532, Abell S0726, Abell 3535, Abell S0733
\end{itemize}
For a more complete analysis we add three objects which have no
measured redshift but which were considered as SC members in previous
studies: Abell $3548$, Abell $3561$ and Abell $3563$~\citep{Ein97}.\\


\section{The detection algorithm}
\label{sec:Algorithm}
Our choice of the detection algorithm arises from the need to
analyze a wide area of the sky and to detect extended and most
probably irregular and interacting nearby sources in a high
density environment. We therefore need a fast running algorithm which
does not make any model assumption on the sources to detect.

Our ad hoc written algorithm directly analyses photon events (which
allows a null loss of resolution). Initially a tree-based binning of
the photon distribution is performed. The aim of the binning is to
subdivide the image into the lowest possible number of cells, each
containing only one photon. The binning avoids the need to calculate
exact distances between all photon pairs in the image; distances
between two photons are computed only in a few special cases. This
strongly reduces computational time, still allowing an easy, fast
and accurate study of the photon positions, keeping the photon event
resolution ($0.5$ arcsec) of the initial raw data.

The construction of the tree of cells is performed applying a
rectangular division of the two-dimensional space into (sub)cells. The
initial image is subdivided into two sub-images, splitting it in the
direction perpendicular to its longer side, in such a way that the
daughter cells have half the number of particles of the parent
image. The two sub-images are then analyzed independently; they are
further split perpendicularly to their longer sides, so that their
daughter cells have half the number of particles. All daughter cells
containing more than one photon are recursively sub-divided into
sub-subcells using the above criterion. The binning ends when all
subcells contain only one photon.

After subdividing the image into a number of rectangles equal to the
number of photons present in the field, the algorithm then searches
for spatial density enhancements in the photon distribution with a
{\it friends-of-friends} approach (a modified version of what made publicly 
available by the University of Washington at:
http://www-hpcc.astro.washington.edu/tools/). 
This is performed by gathering in
the same group all photons which are within a predefined distance
$r_{\rm max}$ of each other. Eventually a group is formed by photons
all having at least one ``friend'' within a distance smaller or equal
than the specified $r_{\rm max}$. Each detected group, formed by at
least $N_{\rm min}$ photons, is considered a candidate cluster. A more
detailed description of the algorithm can be found 
in~\cite{DeF03}.~\footnote{A complete version of the Ph.D. 
thesis can be downloaded from: 
${\rm http://people.na.infn.it/\sim betty/tesi\_files/PhD.pdf.gz.}$}

\subsection{Optimizing the algorithm}
\label{sec:optim_alg}
In a field containing only sky background a source 
detection algorithm should detect nothing. We keep this in mind
when tuning the two parameters of our algorithm: the search radius
$r_{\rm max}$ and $N_{\rm min}$, which are the maximum distance within
which photons are considered to belong to the same group, and the
minimum number of photons a group needs to be considered a candidate
cluster, respectively. We hence simulate background photon
distributions. In each simulated field we randomly distribute a number
of photons equal to the observed average photon number ($\approx 16000$)
in the background of RASS-III fields (each covering 6.4 x 6.4 degrees 
of sky).

When we run our detection algorithm on the simulated background
fields, for high values of the search radius $r_{\rm max}$ and for low
values of the minimum number of photons forming each source -$N_{\rm
min}$-, a high number of sources is detected. On the other hand, if a
very small value is assigned to $r_{\rm max}$ and, at the same time, a
group of photons is considered to be a source only if it has a very
high number of photons (large values of $N_{\rm min}$), the algorithm
retrieves only few sources or, to the extreme, no sources at all. If
both the values of $r_{\rm max}$ and $N_{\rm min}$ are low, only
peaked sources with few photon counts are detected. Larger values of
$r_{\rm max}$ have instead to be chosen if one wants to detect
extended sources. For increasing $r_{\rm max}$, the value of $N_{\rm
min}$ has to increase accordingly, in order to reduce the
number of spurious detections.\\ We choose the couple of values which
allows an average detection of only $0.75$ sources in each simulated
background field:
$$r_{\rm max}^{\rm sim}=230\ {\rm bins}\ \ \ \ \ \ \ N_{\rm min}=17\
{\rm counts}$$ where each bin is $0.5\arcsec$. Real photons are though
not randomly distributed because of the large scale structure present
in each field, and since the exposure time changes from field to field
and is not even constant inside each field.

In order to obtain an unbiased detection, we appropriately weight the
values of the search radius $r_{\rm max}^{\rm sim}$ chosen from the
simulations by the local exposure time and background. To this aim 
additional data (like event rates, aspect and quality of
the data, exposure and background images) are used.

We scale $r_{\rm max}^{\rm sim}$ to the true background level within
each pixel in each field, taking into account the following simple
statistics. Let us consider an area $A$ in which there is a constant
density of events $n=N/A$. The probability that an event occurs in a
smaller circular area $a=\pi r^2$ is equal to $p_{\rm int}=a/A$; the
probability that the event occurs instead outside $a$ is equal to
$p_{\rm ext}=(A-a)/A$. The mean number of events $\langle k\rangle$ inside the
region of area $a$ is given by:
$$\langle k\rangle=N \cdot p_{\rm int}= \frac N A \cdot a=n\cdot a=n\cdot \pi r^2$$
Then, the radius inside which $\langle k \rangle$ events are expected to be found can be written, for the simulated fields, as:
\begin{equation}
r_{\rm max}^{\rm sim}=\sqrt{\frac {\langle k\rangle}{\pi}} \cdot \frac 1 {\sqrt{n_{\rm sim}}}
\label{eq:poi_sim}
\end{equation}
where $n_{\rm sim}$ is the number of background counts per unit area in the simulations.
We then scale $r_{\rm max}$ proportionally with the local background in the following way:
\begin{equation}
r_{\rm max}(x,y)=\sqrt{\frac{\langle k\rangle}{\pi}} \cdot \frac 1 {\sqrt{n_{\rm RASS}(x,y)}}
\label{eq:poi_rass}
\end{equation}
where $n_{RASS}$ is the number of local RASS background counts per unit area.\\
Eqs.~(\ref{eq:poi_sim}) and~(\ref{eq:poi_rass}) can be rewritten as:
\begin{eqnarray}
r_{\rm max}(x,y)&=&r^{\rm sim}_{\rm max}\ \sqrt{\rm bkg.\ ph.\ dens.\ (sim.)}\nonumber \\
&\times &\frac 1 {\sqrt{\rm bkg.\ ph.\ dens.\ (RASS\ III)\ (x,y)}}.
\label{eq:scaling}
\end{eqnarray}
Eq.~(\ref{eq:scaling}) scales the search radius according to the
density of the background: in areas with higher background a
proportionally smaller $r_{\rm max}$ will be applied, and vice-versa.

\subsection{Running the algorithm}
On the $25$ selected RASS-III plates we run the above described source
detection algorithm. Some sky regions are covered by more than one,
overlapping, adjacent fields. In such cases the algorithm is run on
all available fields; multiple detections are eliminated.
Sources located on the border of one of the plates show
a mismatch in the measured count-rates within the different 
plates. In these cases we choose the detection made in the plate where
the source is farther away from the border.

X-ray emission coming from the ICM, at temperatures of typically 
$10^7-10^8\ {\rm K}$, occurs predominantly in the hard band. 
Therefore, in order
to obtain the highest possible detection rate with the lowest
contamination by non-cluster sources, the cluster detection algorithm
is initially run on the photon distribution in the hard ROSAT energy
band ($0.5-2.0\ {\rm keV}$).

For each detected source the following parameters are measured:
background-subtracted count rate, number of photons in the hard band
and in the soft band, hardness ratio, source center weighted for the
varying exposure time inside the source area, total area, radius 
(mean photon distance from the center of the cluster - $\langle r\rangle$)
and an estimate of the source extent (${\rm Ext.}=\langle
r\rangle+1.5\:\sigma$, where $\sigma$ is the radius containing 68\%
of the source photons from the group.)


\section{Cluster detection}
\label{sec:cl_detection}
We apply our detection algorithm to the whole surveyed area, detecting a total
of $579$ sources. A cross-correlation with archival data shows that
these can be classified as: $48$ galaxy clusters, $90$ galaxies, $81$ stars,
$26$ infrared sources (IrS) from the 2 Micron All Sky Survey eXtended sources
catalogue~\citep{Jar00}, $158$ point-like X-ray sources from the ROSAT All-Sky
Survey Bright Source Catalogue (BSC)~\citep{Vog99} and $4$ galaxy groups. The
remaining $172$ detected sources are new X-ray detections.\\

\subsection{Applying further selection criteria}
\label{sec:apply_sel}
Finding clusters in X-rays is not an easy task, mainly due to their
paucity compared to other types of sources. A large fraction of the
sources detected by applying our detection algorithm to the whole area
(see \S~\ref{sec:cl_detection}) is in fact given by non-cluster
sources. Since the main aim of this work is to detect possible new 
cluster candidates in the area, we apply further selection criteria to 
all detected sources, with the purpose to exclude the largest possible
number of non-cluster sources and spurious detections. We will then be
able to perform follow-up observations only of sources that have a 
high probability of being cluster candidates.

\subsection{The hardness ratio}
The first selection criterion that we apply is on the spectral hardness of the X-ray emission of each source. Here we use the RASS definition of a source's hardness ratio (HR):
$$HR=\frac{h-s}{h+s}$$
where $h$ is the photon count in the hard energy range from $0.5$ to $2.0\ {\rm keV}$ (energy channels $52-201$), and $s$ is the photon count in the soft band, from $0.1$ to $0.4\ {\rm keV}$ (energy channels $11-40$).

Galaxy clusters have a harder spectrum than other more common X-ray
emitting sources. Therefore in principle most clusters are more easily
detected in a hard X-ray energy band, where the X-ray background is
lower and soft sources are fainter. Unfortunately the ROSAT hard
energy band is extremely limited ($0.5-2.0\ {\rm keV}$), and at a flux
of $\approx10^{-14}\ {\rm ergs\ s}^{-1}\ {\rm cm}^{-2}$ clusters still
comprise only $10\%-20\%$ of the total source
population~\citep{Has93}.

\begin{figure}[ht]
\centering
\resizebox{\hsize}{!}{\includegraphics{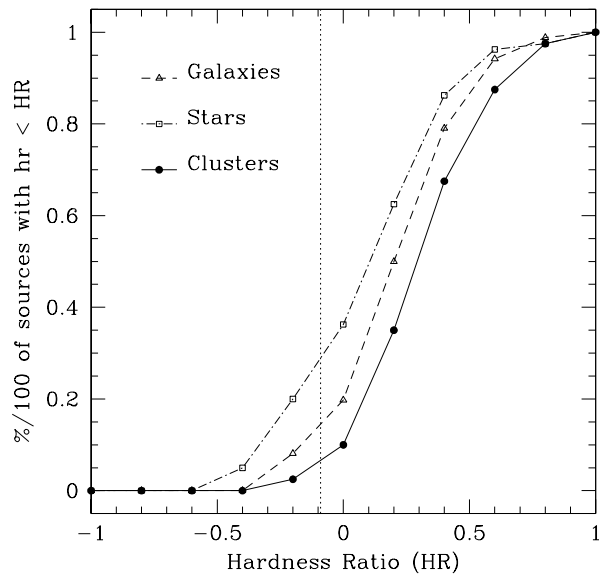}}
\caption{Cumulative normalized distribution of the hardness ratio of a selection of the known clusters, galaxies and stars among the detected sources.}
\label{fig:hr}
\end{figure}

In Fig.~\ref{fig:hr} we show the HR distributions of the
subset of our detections, for which optical counterparts were
found. The cluster HR distribution (plotted as a solid line) is
significantly different from the one for galaxies (dashed
line); the distribution for stars (dot-dashed line) lies between the two. 
A Kolmogorov-Smirnov (KS) test is used to compare the star and the
galaxy HR distributions to the cluster one. The probability that the
galaxy (star) and the cluster distributions are the same is indeed
very low: ${\rm P}_{\rm KS}({\rm Clusters\ vs\ galaxies})=8.5\times
10^{-4}$ (${\rm P}_{\rm KS}({\rm Clusters\ vs\ stars})=8.2\times
10^{-3}$). There is however no threshold in the HR
value which allows a clean
discrimination between clusters and other types of sources.
The value which provides the best compromise, by allowing us to eliminate
the highest number of soft sources while losing the lowest possible 
number of clusters is: 
$${\rm HR}=-0.09.$$
which corresponds to a temperature of $kT\approx 0.5\ {\rm keV}$.

After applying the above HR selection criterion to all detected sources, we are left with $413$ sources among which there are: $40$ galaxy clusters, $25$ galaxies, $76$ stars, $22$ infrared sources, $120$ point-like X-ray sources and $4$ galaxy groups. The remaining $126$ sources are new X-ray detections.\\
\label{sec:HR}

\subsection{The source extent}
\label{sec:extent}
Clusters of galaxies are much more extended structures than single galaxies or stars. If all sources were at the same distance from us a threshold value in the minimum/maximum extent of a source would be an extremely efficient way to discriminate between extended and point sources. If, however, one is interested in objects at various distances, up to very high redshifts, the source dimension is not such a straightforward selection criterion, and extra care has to be taken in choosing a threshold value. Since we are interested exclusively in low redshift objects, a selection based on the source extent can still be a valid source discriminator.

\begin{figure}
\centering
\resizebox{\hsize}{!}{\includegraphics{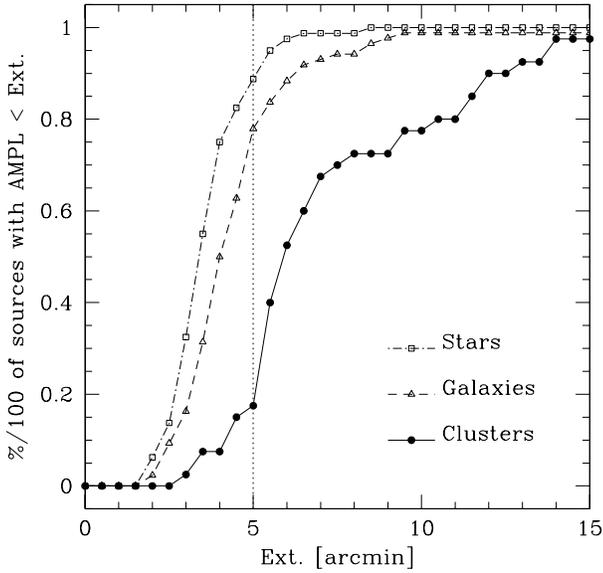}}
\caption{Cumulative normalized distributions of the spatial extent of a sample of the known stars, galaxies and clusters among all detected sources.}
\label{fig:ampl}
\end{figure}

Figure~\ref{fig:ampl} shows the cumulative distributions of the radial
spatial extent (Ext.) for a sub-sample of our detections, for
which optical counterparts were found. At low values of Ext., the cluster
distribution (plotted as a solid line) is much flatter than the galaxy
(dashed line) and the star (dot-dashed line) ones, which show instead
a much steeper increase even for low radii. The value which allows the
best discrimination
between clusters and less extended sources is:
$${\rm Ext.}=300\arcsec,$$
where the source extent has been defined in \S~\ref{sec:optim_alg}.

A KS test is applied to the star and the galaxy distributions to
compare them with the cluster one. According to this test the two
distributions are indeed very different:
${\rm P}_{\rm KS}({\rm Clusters\ vs\
galaxies})=1.5\times 10^{-9}$ and ${\rm P}_{\rm KS}({\rm Clusters\ vs\
stars})=7.8\times 10^{-13}$.

After applying the above threshold to the source extent of all
remaining detected sources we are left with $141$ objects. Among these
there are $34$ galaxy clusters, $25$ galaxies, $15$ stars, $8$
infrared sources, $21$ point-like X-ray sources,
$4$ galaxy groups and $34$ previously unknown sources.\\

\subsection{Minimum count rate}
In order to have a flux lower limit, for all the sources detected in
the area, independent of the different exposure times across the
field, a count rate threshold of $0.07\ {\rm counts\ s}^{-1}$ in the
$0.5-2.0\ {\rm keV}$ energy band is applied to all the remaining
detections; this corresponds approximately to a cut in flux of $f_{\rm
X}(0.1-2.4\ {\rm keV})=1.55\times 10^{-12}\ {\rm ergs\ cm}^{-2}\ {\rm
s}^{-1}$. This last selection leaves a total of $102$ X-ray detections
in the whole area.


\section{Clusters detected in the Shapley region}
\label{sec:fin_res}
After applying our selection criteria to all detected sources, we are left with $102$ objects:
\begin{itemize}
\item $34$ previously known galaxy clusters of which;
\begin{itemize}
\item $9$ with no previous X-ray detection;
\end{itemize}
\item $23$ galaxies;
\item $8$ stars;
\item $7$ infrared sources;
\item $12$ point-like X-ray sources from the BSC;
\item $4$ galaxy groups;
\item $14$ new detections: X-ray extended sources for which no counterpart 
in other wavelengths could be found.
\end{itemize}
We plan to further investigate the $12$ BSC X-ray sources optically, 
since we detect them as extended structures, and hence can be potential clusters.

\begin{figure}
\centering
\resizebox{\hsize}{!}{\includegraphics{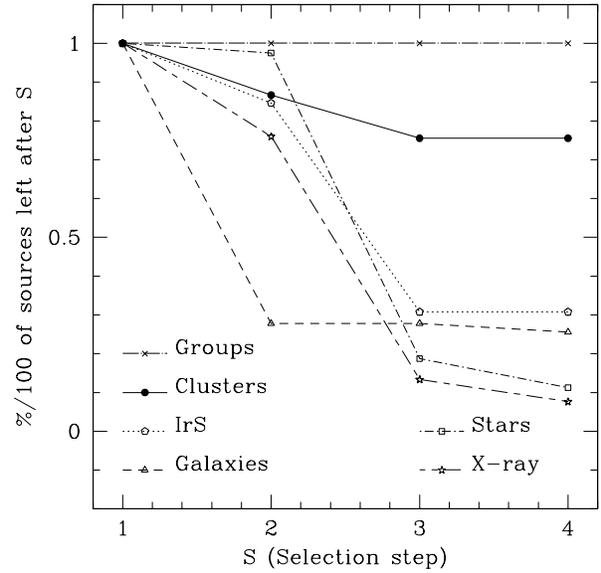}}
\caption{Percentage of different types of detected sources after each step of our detection and selection process.}
\label{fig:selection}
\end{figure}

In Fig.~\ref{fig:selection} we plot the percentage of the different
types of sources detected which remain after applying our 
selection criteria:
(1) detection algorithm, (2) HR, (3) extent and (4) minimum flux selection
criteria were applied. It is clear that we are left with a high
percentage of diffuse sources (i.e. clusters and groups of galaxies),
as opposed to more point-like and softer sources of which 
most were rejected.
As expected, having applied ad hoc selection criteria
to all detected sources has largely improved the performance of the
detection algorithm, leading to the rejection of a reasonably 
low number of clusters ($15$) and not even one galaxy group, while still
performing a good selection against spurious detections and different
types of sources.

Here follows a list of the $34$ detected clusters in the area with (\ref{point:with_prev}.) and without (\ref{point:without_prev}.) previous X-ray detection:
\begin{enumerate}
\item Abell S0700, Abell S0701, Centaurus, RX J1252.5-3116, Abell
3528, Abell 3530, Abell 1644, Abell 3532, Abell 1648, Abell 3541,
Abell 3548, Abell 1732, Abell 1736, Abell 3558, SC 1327-312, SC
1329-314, RX J1332.2-3303, Abell 3562, Abell 1757, Abell 3563, Abell
3565, Abell 3570, Abell 3571, Abell 3574, Abell 3581
\label{point:with_prev}
\item Abell S0717, Abell S0721, Abell S0753, Abell 1604, Abell 1633, Abell 1791, Abell 3519, Abell 3535, Abell 3557 \label{point:without_prev}
\end{enumerate}

Seventeen of these clusters are in the selected redshift range ($0.03
\leq z \leq 0.07$, see \S~\ref{sec:known_cl}); unless otherwise
stated, from now on we will limit our analysis to clusters within this
redshift range. In part of our following analysis we will also
consider the $14$ new X-ray detections, which also are hard, extended
sources located in the area, and are therefore probable SC members.

Eight clusters inside the selected redshift range, with known optical
counterparts, are identified by the detection algorithm
but excluded during one of the three selection steps. These clusters
are:
\begin{itemize}
\item Abell S0758, Abell 1631, Abell 3554, Abell S0718, Abell 3556, Abell 3553, Abell S0724, Abell S0729
\end{itemize}
Four of them are excluded because of their relatively low extent; the
remaining four clusters have instead extremely low values of the
HR. These sources are known to be clusters, because of their optical
counterparts, and are all above our flux limit; we therefore include
them in our following analysis of the SC since they were detected with
the same technique as for all other clusters.

\subsection{Cluster properties}
\subsubsection{Average temperature}
\label{sec:cl_temp}
For all detected objects we estimate the average gas temperature applying the following method by~\cite{Ebe96}.
\begin{enumerate}
\item The flux in the energy band $0.1-2.4\ {\rm keV}$ is initially measured assuming a cluster temperature of $5.0\ {\rm keV}$;
\item The corresponding luminosity is computed;
\item The temperature is derived from the X-ray luminosity through~\citep{Mar98a}:
\begin{equation}
kT=6\ \left[{\rm keV}\right] \left( \frac{L_{\rm X}\left[10^{44}\ {\rm ergs}\ {\rm s}^{-1}\right]}{1.41\ h^{-2}}\  \right)^{\left(\frac{1}{2.10}\right)};
\label{eq:temp}
\end{equation}
\item The estimated temperature is used to recompute the flux.
\end{enumerate}
Steps from (2) to (4) are repeated iteratively until the results converge, which most often happens after no more than three iterations. 

Table~\ref{tab:ris_fin} summarizes measured and estimated quantities
for detected clusters and for the new X-ray detections. For some of
these clusters (Abell 3535, Abell S0718, Abell S0721, Abell S0724,
Abell 3553, Abell 3554, Abell S0729, Abell S0758), and naturally for
all the new detections, this is their first X-ray detection; values in
Table~\ref{tab:ris_fin} are therefore today the only X-ray data
available for these clusters. When available the literature, 
we quote average gas
temperatures obtained from spectral analysis;
the respective references are also listed. We observe a good agreement
between our temperature estimates and values measured in literature.

\subsubsection{X-ray versus optical position}
\begin{figure}
\centering
\resizebox{\hsize}{!}{\includegraphics{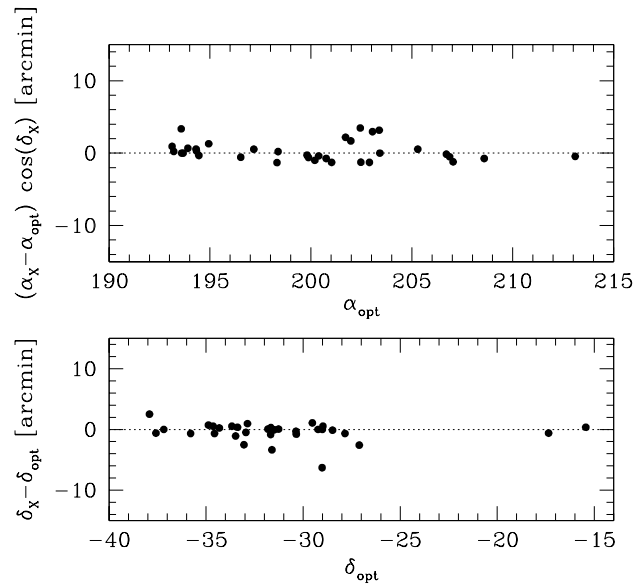}}
\caption{Comparisons, for the detected clusters, of their optical
coordinates with their X-ray detection coordinates. RA and Dec are
separately analyzed to check for systematic deviations.}
\label{fig:confr_posiz}
\end{figure}
In Fig.~\ref{fig:confr_posiz} we compare the optical coordinates of
all detected clusters with the center of our X-ray detection. We
observe a general agreement between their centers in the two wavebands
with no systematic deviation. The few exceptions are represented by
extraordinarily extended clusters or clusters showing
significant sub-structures in their X-ray emission. Even if the peak in the
optical luminosity does not always coincide with the X-ray one, it
falls anyway within the X-ray extent of the cluster.

\subsection{Detection efficiency}
\label{sec:det_eff}
The main aim of this work is to detect the largest possible number of extended
structures that lie in the Shapley SC. We stress that we do not aim to build a
complete sample, but mainly to detect new possible Shapley members in order to
provide a more reliable mass estimate and morphological description of the
whole super-structure. 
In this section we briefly discuss the
detection efficiency of our algorithm by considering the performance on
known galaxy-clusters at low redshift.
Within the surveyed area there are 52 optically known galaxy clusters
with redshift $z \leq 0.07$, of which 47 belong to the Shapley
SC. 
To this sample we add in our studies three more candidates with no redshift estimate
which have been considered as SC members in the past (\S~\ref{sec:known_cl}).
We detect $56\%$ (30) of these 55 clusters applying our algorithm, as outlined
in \S~\ref{sec:cl_detection}. While we miss a significant fraction of real
clusters with our conservative selection criteria, we substantially
reduce the contamination of our sample with non-cluster sources. In our final
catalog of 102 objects we have 38 known galaxy groups/clusters, 50 point
sources and 14 new X-ray detections. Regarding only the known sources we can
expect that about $40\%$ of the new detections are hitherto
unknown galaxy groups/clusters. 

For each optically known cluster which we do not detect with the
blind run of our algorithm we perform a refined second step 
analysis (see \S~\ref{sec:non_detected}).

\subsection{Second step analysis}
\label{sec:non_detected}
Among the optically known galaxy clusters in the area, within the
selected redshift range, there are $25$ clusters which were not
detected by our algorithm. Only four of these $25$ clusters (Abell 1709, Abell 3559, Abell 3560 and Abell 3566) have previous X-ray detections. Their cases are discussed in detail in the following sections.

For these $25$ sources we perform a
further analysis to verify if any X-ray emission is
present at their location. We run our detection algorithm lowering the
minimum number of photons allowed in a group, and applying no further
cuts. Despite the low threshold we can not detect any
X-ray emission for $10$ of the $25$ clusters:
\begin{itemize}
\item Abell 3544, Abell 3555, Abell S0731, Abell S0733, Abell 3559, Abell 3560, Abell 3561, SC 1342-302, Abell S0740, Abell 3578.
\end{itemize}
For the remaining $15$ clusters:
\begin{itemize}
\item Abell 3537, Abell 3542, Abell S0726, Abell 1709, Abell 3552, Abell S0734, Abell 3564, SC 1336-314, Abell 3566, Abell 3568, Abell S0742, Abell 3572, Abell 3575, Abell 3577, Abell S0757
\end{itemize}
we instead detect signs of faint X-ray emission. Properties of these {\it second step} detections are listed in Table~\ref{tab:undetected}, while in Table~\ref{tab:scheme} we provide the number of optically known and detected clusters located in the redshift ranges and sky regions analyzed in this paper.

\begin{table*}
\caption{Sources detected in the second step analysis}             
\label{tab:undetected}      
\centering          
\begin{tabular}{lccccrccccc}
\hline\hline       
Cluster & RA; Dec  & z &$n_{\rm H}$& Radius & HR & Count Rate & $f_{\rm X}$ & $f_{\rm bol}$ & $L_{\rm bol}$ & $kT$\\ 
\cline{8-9} &(J2000.0) & &$(10^{20}$ & & & & \multicolumn{2}{c}{$(10^{-12}$} &$(10^{44}$ &\\ 
& (deg) &  &${\rm cm}^{-2})$ &
(arcmin) &  & $\left( {\rm counts\ s}^{-1}\right)$ & \multicolumn{2}{c}{${\rm ergs\ cm}^{-2}\ {\rm s}^{-1})$} &${\rm ergs\ s}^{-1})$ & (keV)\\ 
\hline  
Abell 3537 &$195.255;-32.423$ &$0.0320$ &$5.70$ &$2.80$ &$+0.47$
&$0.034$ &$0.85$ &$1.13$ &$0.070$ &$0.92$\\ Abell 3542 &$197.201;-34.489$
&$0.0525$ &$4.82$ &$1.82$ &$+0.00$ &$0.017$ &$0.46$ &$0.65$ &$0.040$ &$0.69$\\
Abell S0726 &$198.806;-33.695$ &$0.0590$ &$4.78$ &$2.79$ &$+0.43$ &$0.026$
&$0.55$ &$0.85$ &$0.069$ &$0.90$\\ Abell 1709 &$199.690;-21.488$ &$0.0521$
&$8.01$ &$5.16$ &$+0.22$ &$0.052$ &$1.30$ &$1.73$ &$0.106$ &$1.12$\\ Abell
3552 &$199.753;-31.851$ &$0.0520$ &$4.72$ &$1.48$ &$+0.20$ &$0.018$ &$0.49$
&$0.70$ &$0.041$ &$0.70$\\ Abell S0734 &$202.006;-41.106$ &$0.0503$ &$6.75$
&$1.72$ &$+0.67$ &$0.017$ &$0.51$ &$0.71$ &$0.040$ &$0.67$\\ Abell 3564
&$203.660;-35.228$ &$0.0505$ &$4.08$ &$1.14$ &$+0.43$ &$0.016$ &$0.46$ &$0.62$
&$0.035$ &$0.66$\\ SC 1336-314 &$204.071;-31.788$ &$0.0395$ &$3.82$ &$2.47$
&$+0.20$ &$0.025$ &$0.72$ &$0.98$ &$0.034$ &$0.64$\\ Abell 3566
&$204.762;-35.610$ &$0.0510$ &$4.16$ &$1.04$ &$+0.75$ &$0.021$ &$0.56$ &$0.76$
&$0.044$ &$0.72$\\ Abell 3568 &$205.328;-34.652$ &$0.0516$ &$4.10$ &$1.67$
&$+0.43$ &$0.016$ &$0.46$ &$0.62$ &$0.036$ &$0.66$\\ Abell S0742
&$206.107;-34.305$ &$0.0510$ &$4.41$ &$1.57$ &$+0.71$ &$0.018$ &$0.50$ &$0.69$
&$0.040$ &$0.68$\\ Abell 3572 &$207.100;-33.394$ &$0.0517$ &$4.13$ &$2.22$
&$+0.39$ &$0.026$ &$0.66$ &$0.89$ &$0.052$ &$0.79$\\ Abell 3575
&$208.170;-32.920$ &$0.0377$ &$4.50$ &$2.42$ &$+0.25$ &$0.030$ &$0.85$ &$1.17$
&$0.037$ &$0.66$\\ Abell 3577 &$208.574;-27.895$ &$0.0494$ &$4.67$ &$1.29$
&$+1.00$ &$0.020$ &$0.55$ &$0.75$ &$0.041$ &$0.70$\\ Abell S0757
&$213.035;-33.086$ &$0.0440$ &$4.76$ &$2.17$ &$-0.20$ &$0.018$ &$0.55$ &$0.74$
&$0.032$ &$0.63$\\
\hline                  
\end{tabular}
\begin{list}{}{}
\item Col. (1): Source name. Col. (2): RA and Dec(J2000). Col. (3): Redshift. Col. (4): Galactic HI column density in units of $\left(10^{20}\ cm^{-2}\right)$~\citep{Dic90}. Col. (5): Source radius (arcmin). Col. (6): Hardness Ratio. Col. (7): Count Rate in the $0.5-2.0\ {\rm keV}$ energy band. Cols. (8) - (9): $f_{\rm X}$ in the $0.1-2.4\ {\rm keV}$ energy band and bolometric flux both in units of $\left(10^{-12}\ {\rm ergs\ cm}^{-2}\ {\rm s}^{-1}\right)$. Col. (10): Bolometric luminosity in units of $\left(10^{44}\ {\rm ergs\ s}^{-1}\right)$. Col. (11): Cluster average temperature in keV.
\end{list}
\end{table*}

\begin{table*}
\caption{Optically known and detected clusters}             
\label{tab:scheme}
\centering          
\begin{tabular}{l|l|c|c|c|c}
\hline\hline
 & & $z<0.03$  & $0.03\leq z \leq 0.07$ & $0.0446\leq z \leq 0.0554$ & $0.0446\leq z \leq 0.0554$\\ 
 & &   &  &  & $208.74\leq {\rm RA} \leq 193.14$\\
 & &   &  &  & $-37.64 \leq {\rm Dec}\leq -27.15$\\
\hline\hline
$1$ &N$_{\rm cl}$  & $5$  & $46\ (+1)$ & $34\ (+1)$& $30\ (+1)$\\
\hline
$2$ &Detected  & $5$  & $17$ & $14$& $11$\\
$3$ &Detected but excl.  & $-$  & $8$ & $7$& $6$\\
$4$ &Det. in $2^{\rm nd}$ step  & $-$  & $15$ & $10$& $8$\\
\hline                  
$5$ &TOT detected  & $5$  & $40$ & $31$& $25$\\
\hline                  
\end{tabular}
\begin{list}{}{}
\item Col. (1): Line number. Col. (2): Table legend. Number of clusters optically known (line 1), detected by our algorithm (line 2), identified by the detection algorithm but excluded during one of the three selection steps (line 3), detected during the second step analysis (line 4); in line 5 is the total number of clusters detected in this work (= line 2 + 3 + 4). The number of clusters are listed in different redshift ranges. Col. (3): $z<0.03$. Col. (4): $0.03\leq z \leq 0.07$. Col. (5): within the redshift range of the core of the Shapley SC ($0.0446\leq z \leq 0.0554$). Col. (6): in the sky region of the core of the Shapley SC. In line 1 we list RX J1332.2-330 in parenthesis since from our analysis shows up as a probable coincidence with Abell~3560 (see \S~\ref{sec:3560}).
 \end{list}
\end{table*}

\subsubsection{Abell 1709}
The only previous X-ray detection of Abell 1709 is by~\citet{Jon99}
from pointed observations with the Einstein satellite. Within a radius
of $1\ {\rm Mpc}$ they measure a cluster X-ray luminosity of $L_{\rm
X}(0.5-4.5\ {\rm keV})=0.176\times 10^{44}\ {\rm ergs\ s}^{-1}$
(transformed for our assumed cosmology); this corresponds to a
bolometric luminosity of $L_{\rm X} ({\rm bol})=0.296\times 10^{44}\
{\rm ergs\ s}^{-1}$ and to a flux of $f_{\rm X}(0.1-2.4\ {\rm
keV})=3.63\times 10^{-12}\ {\rm ergs\ cm}^{2}\ {\rm s}^{-1}$. At the
local exposure time of $t_{\rm exp}=256\ {\rm s}$ the cluster should
show a total of $37$ photons. We instead detect only $15$ photons for
this source. Because of the high background and the faint signal from
the source, we can only trace the cluster out to a radius of
$0.3\ {\rm Mpc}$; this might partially cause the observed flux
discrepancy.

\subsubsection{Abell 3559}
The only previous X-ray detection of Abell 3559 is by~\citet{Dav99}
from pointed ROSAT observations. Within a radius of $1\ {\rm Mpc}$
they estimate a bolometric luminosity for the cluster of $L_{\rm
X}({\rm bol})=0.226\times 10^{44}\ {\rm ergs\ s}^{-1}$, which
corresponds to a flux of $f_{\rm X}(0.1-2.4\ {\rm keV})=3.4\times
10^{-12}\ {\rm ergs\ cm}^{-2}\ {\rm s}^{-1}$. At the local exposure
time of $t_{\rm exp}=294\ {\rm s}$ this would correspond to a total of
$42$ cluster photons. This is much higher than our flux limit and
hence we would expect to detect this source, but neither the hard nor
the soft RASS-III images show any increase above the background
emission at the cluster position. Deeper X-ray observations of this object 
are needed to clarify this discrepancy.

\subsubsection{Abell 3560}
\label{sec:3560}
\begin{figure}
\centering
\resizebox{\hsize}{!}{\includegraphics{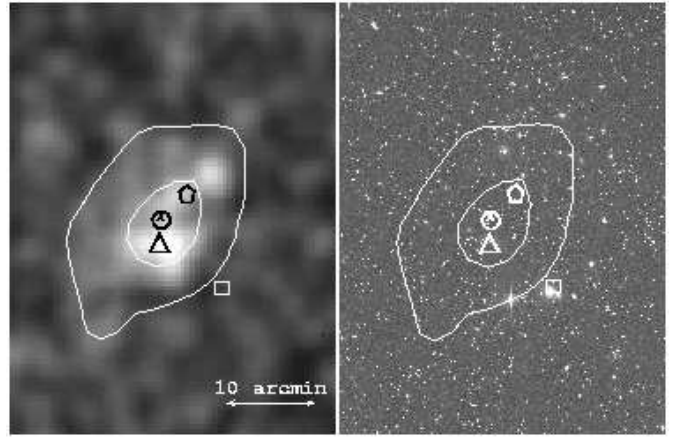}}
\caption{Abell 3560. {\bf Left}: Smoothed X-ray image with
superimposed X-ray contours at $0.2$ and $0.4\ {\rm counts\ px}^{-1}$. 
The empty square and pentagon point to the positions of
Abell 3560 and RXJ 1332.2-3303; the cross is for the location of the
detection of RXJ 1332.2-3303 in this work and the circle and triangle
point to the locations of the detections of Abell 3560 by~\cite{Ebe96}
and~\cite{Dav99}. {\bf Right}: optical DSS image.}
\label{fig:3560}
\end{figure}

Abell 3560 was already detected in X-rays by~\citet{Ebe96} using
RASS-I data and by~\citet{Ett97} and by~\citet{Dav99} through ROSAT
pointed observations. Their values of the cluster bolometric
luminosities, converted to our cosmology, are: $L_{\rm X}({\rm
bol})=1.18\times 10^{44}\ {\rm ergs\ s}^{-1}$, $L_{\rm X}({\rm
bol})=1.16\times 10^{44}\ {\rm ergs\ s}^{-1}$ and $L_{\rm X}({\rm
bol})=1.40\times 10^{44}\ {\rm ergs\ s}^{-1}$, respectively.

In the left panel of Fig.~\ref{fig:3560} is the RASS hard X-ray image
of A~3560. A bright, extended and irregular source is observed close
to the position of RX J1332.2-3303 (pentagon), while there is no
detectable X-ray emission at the location of Abell~3560 (empty box).
The centers of the X-ray detections of both~\cite{Ebe96}
and~\cite{Dav99} (plotted as an empty circle and triangle,
respectively) lie close to the position in the sky of RX J1332.2-3303
and well within the extent of our X-ray detection; the distance to
Abell~3560 is instead much larger [${\rm d}_{\rm RX}=4.2 \arcmin$ and
${\rm d}_{\rm Ab}=11.5 \arcmin$ for~\citet{Ebe96} and ${\rm d}_{\rm
RX}=6.5 \arcmin$ and ${\rm d}_{\rm Ab}=10.2 \arcmin$
for~\citet{Dav99}]. Our X-ray detection of RX J1332.2-330 (cross) is
coincident with the detection of Abell~3560
by~\cite{Ebe96}. \citet{Ett97} do not explicitly give coordinates for
their detection; no comparison with their work was therefore possible.

We are therefore inclined to think that either the X-ray emission of Abell 3560 is too faint to be observable even in pointed observations, or more probably that Abell 3560 and RX J1332.2-3303 are the same cluster and that the coordinate mismatch is simply due to the fact that the optical position was centered on a bright galaxy which lies on the outer edges of the cluster.

Assuming the latter hypothesis to be true, we recompute our
bolometric luminosity for RX J1332.2-3303 using the value of the
temperature ($kT=3.4\ {\rm keV}$) measured by~\cite{Dav99} and the
redshift of Abell 3560; we obtain a value $L_{\rm X}({\rm
bol})=1.13\times 10^{44}\ {\rm ergs\ s}^{-1}$ which is only slightly
lower than those computed by~\citet{Ebe96}, \citet{Ett97}
and~\citet{Dav99}; we see the same trend
in our estimate of the cluster temperature (see Table~\ref{tab:ris_fin}).

\subsubsection{Abell 3566}
As for Abell 3559, also for Abell 3566 one previous detection in
X-rays is present in the literature (by~\cite{Dav99} using ROSAT pointed
observations). Within $1\ {\rm Mpc}$ they compute a cluster luminosity
of $L_{\rm X}(0.5-2.0\ {\rm keV})=0.029\times 10^{44}\ {\rm ergs\
s}^{-1}$; this corresponds to a bolometric luminosity of $L_{\rm
X}({\rm bol})=0.086\times 10^{44}\ {\rm ergs\ s}^{-1}$ and to a flux
of $f_{\rm X}(0.1-2.4\ {\rm keV})=1.10\times 10^{-12}\ {\rm ergs\
cm}^{-2}\ {\rm s}^{-1}$, which is below our lower flux
limit. Furthermore, in the hard band, where we initially detect the
sources, the cluster would have a count rate of: ${\rm cr}(0.5-2.0\
{\rm keV})= 4.2\times 10^{-2}\ {\rm counts\ s}^{-1}$, which at the
local exposure time of $t_{\rm exp}=323\ {\rm s}$ gives a total number
of photon counts for the whole source of $N_{\rm tot}=13$, which is
below the minimum limit fixed in the algorithm (see
\S~\ref{sec:optim_alg}). We therefore detect this cluster only in the
second step of our analysis. From the RASS data the cluster is barely
detectable and appears almost point-like, with a radius of $\sim 100\
{\rm kpc}$. We therefore detect only a small percentage of its
emission, which causes an even lower detected flux of: $f_{\rm
X}(0.1-2.4\ {\rm keV})=0.56\times 10^{-12}\ {\rm ergs\ cm}^{-2}\ {\rm
s}^{-1}$.


\section{New X-ray detections}
\label{sec:new_det}
Besides already known groups and clusters of galaxies, we also
detect $14$ new cluster candidates, which also are hard, extended
X-ray sources above our flux limit, but have no previous X-ray
detection and no known optical counterpart. Details of the X-ray
properties of these $14$ sources (labeled as B1-B14) are given in 
Table~\ref{tab:ris_fin}.

To verify these cluster candidates in the optical we have performed follow-up 
observations.

\subsection{Optical follow-up}
The primary aim of this follow-up program is the identification of
optical counter parts for the X-ray peaks. In particular we want to
closer investigate whether these sources are hitherto unidentified
members of the Shapley SC. To this end we try to identify galaxy
over-densities and to look for Red Cluster sequences from elliptical
cluster galaxies at $z\approx 0.05$. We obtained optical imaging data
for 11 out of the 14 fields~\footnote{No observations or data of very
bad quality were obtained for fields B2, B3 and B10.} with WFI@ESO2.2m
having a field-of-view of 34\arcmin 0$\times$34\arcmin 0 (ESO
Observing Programs 71.A-0430(A,B); P.I.  S. Schindler). For each
candidate we acquired 2350 s (divided in 5 exposures a 470 s)
in broad-band $R$ and 2010 s (divided in 3 exposures a 670 s) in
broad-band $B$. The fields were observed by ESO in queue scheduling
from April 2003 to September 2003 mostly under non-favorable
photometric conditions. The processing of the data was done with the
image processing pipeline and the methods described in~\cite{Sch03,Esd05} 
and the details are not repeated here. The seeing in the final
stacked mosaic images ranges from 1.0 to 1.8 arcsec and because of the
lack of usable standard star observations we estimated approximate
magnitude zero-points by matching the galaxy number counts to WFI
observations of the Chandra Deep Field South~\citep{Gia04}.

We give a detailed discussion of our findings for each candidate in
App.~\ref{app:properties}.  Our main result is that we consider the
cases B1, B4, B5 and B6 as good candidates for new galaxy clusters. In
these cases, we see a clear light over-density from galaxies close to
the X-ray emission peaks and have indications for a Red Cluster
sequence in the color-magnitude plots.  In the course of our work we
noticed that our candidate B5 corresponds to the Abell cluster
3538. Only further observations (e.g. spectroscopy) can finally
clarify the nature of these sources. If they indeed are members of the
Shapley SC they all represent, as expected, optically poor galaxy
cluster systems (\S~\ref{sec:XLF}). 
For the cases B7, B9, B11, B12 and B13 the most probable explanation
for the X-ray emission at this stage are point sources and we do not
think that the X-ray flux in these fields originates from Shapley SC
members. No conclusive results are obtained in the cases B8 and
B14. In B8, probably most of the X-ray emission comes from a galaxy
group at $z\approx 0.01$ but we also see a slight light over-density at
the X-ray peak position.  This candidate is included in the REFLEX
galaxy cluster catalog~\citep{bsg04}.  B14 is located in a dense stellar field
but shows an extended light over-density at the X-ray peak.


\section{Cluster number density}
\label{sec:morph_anal}
\begin{figure}
\centering
\resizebox{\hsize}{!}{\includegraphics{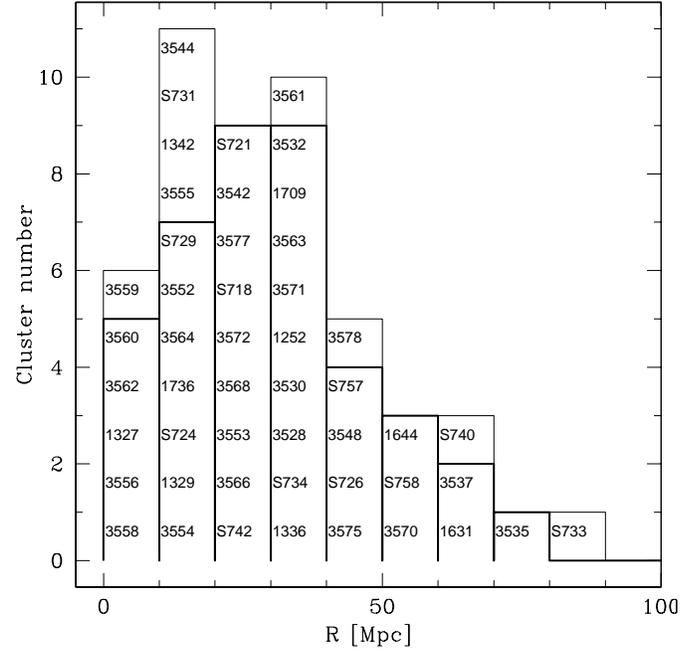}}
\caption{Histogram of the number of clusters as a function of their co-moving distance from the cluster Abell 3558. Over-plotted with a bold line is the histogram of the clusters detected in this work.}
\label{fig:isto_cl}
\end{figure}
Figure~\ref{fig:isto_cl} shows a histogram of the number of clusters
in the region, within the $0.03\leq z \leq 0.07$ redshift range, 
as a function of their co-moving separation from
Abell 3558; we choose this cluster as a reference point since it is one
of the most luminous and therefore most massive clusters in the whole
area and it is located at the center of the densest region of the
super-cluster. The co-moving distance between two clusters is defined
as:
$$R_{\rm A3558,i}=\sqrt{D^2_{\rm A3558}+D^2_{\rm i}-2D_{\rm
A3558}D_{\rm i}\cos \phi}$$ where $D_{\rm A3558}$ and $D_{\rm i}$ are
the co-moving distances to Abell 3558 and to the cluster $i$,
respectively, while $\phi$ is the angular separation between Abell
3558 and the cluster $i$. Over-plotted with a bold line is the
histogram of the clusters which have been detected in this work.\\ Two
peaks in the cluster distribution can be seen; they are due to the
presence of two dense cluster concentrations. The first, more massive
one, is centered at RA$=202.35$, Dec$=-31.88$ and lies around Abell
3558; the second, smaller concentration, is centered at RA$=194.26$,
Dec$=-30.04$ and is at a distance of $\approx 35\ {\rm Mpc}$ from
Abell 3558.

\begin{figure}
\centering
\resizebox{\hsize}{!}{\includegraphics{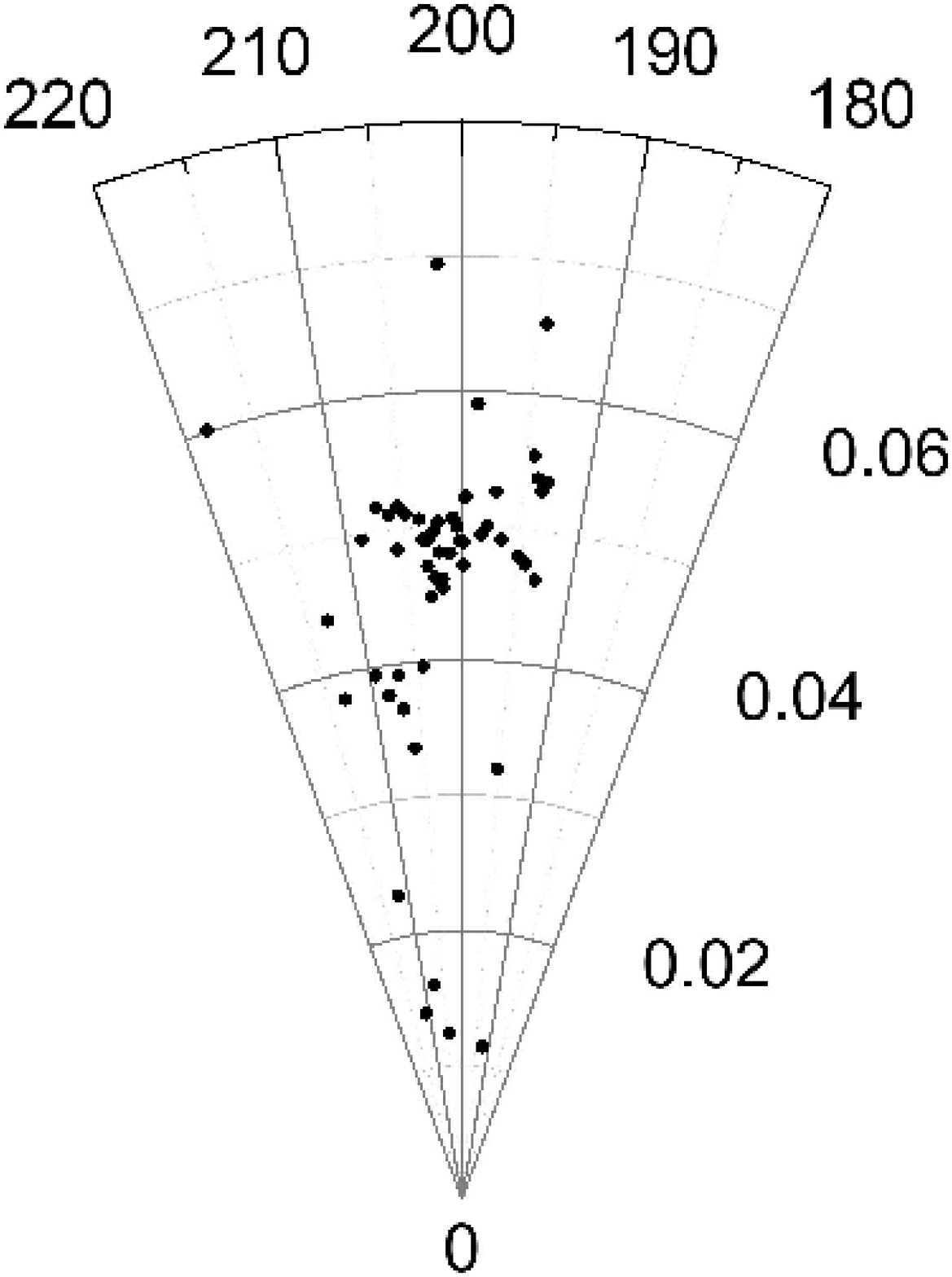}\includegraphics{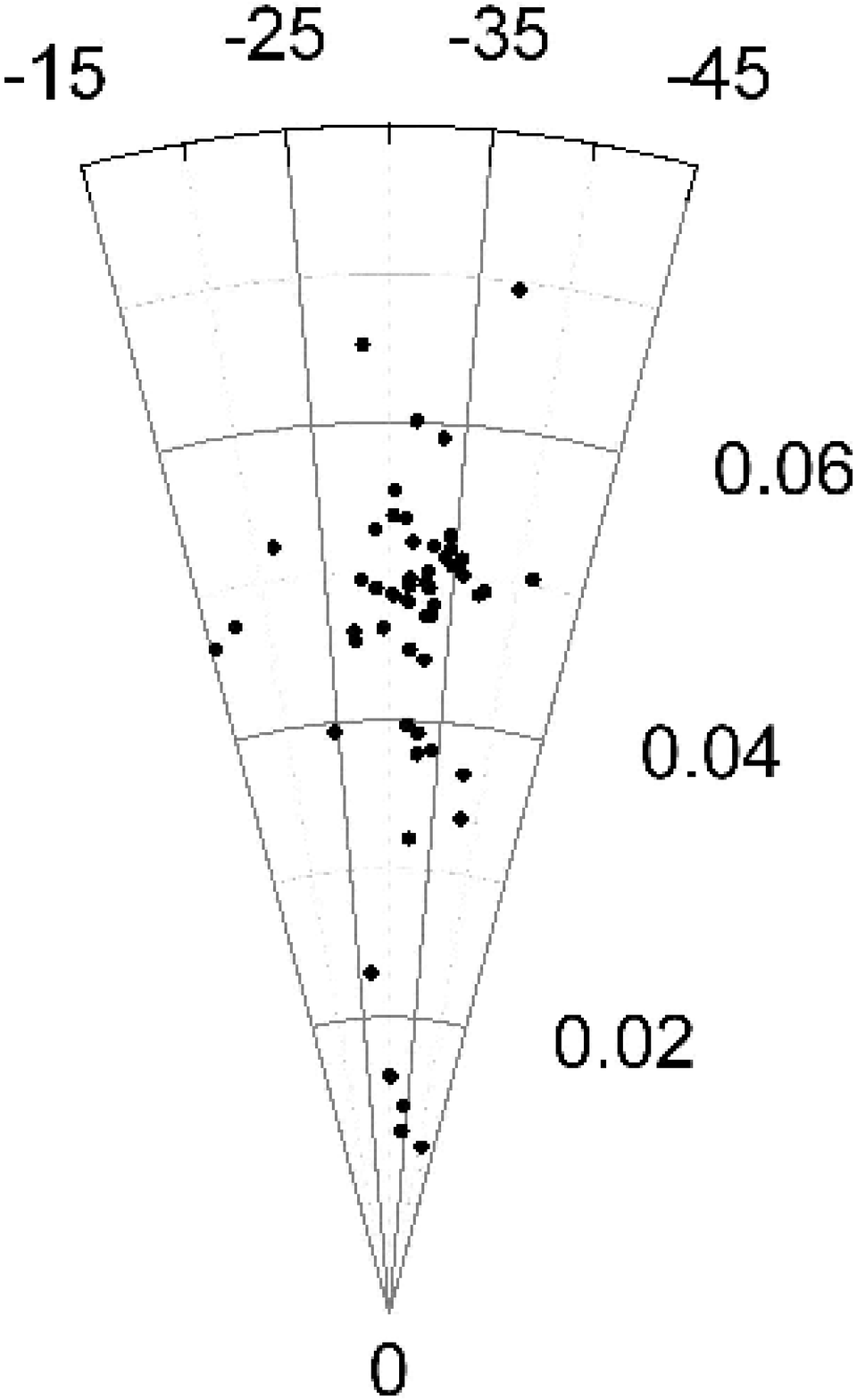}}
\caption{{\bf Left} and {\bf  right panels}: Right Ascension and Declination versus redshift for all the clusters inside the selected redshift range.}
\label{fig:cl_distr_tot}
\end{figure}
The two panels in Fig.~\ref{fig:cl_distr_tot} show the distribution of
all known clusters in the
selected area, which have a measured redshift lower than $0.08$. 
The Shapley SC clearly appears as a dense and compact
concentration of clusters with the densest core lying within redshifts
$z=0.044$ to $z=0.055$. A smaller aggregate of $8$ clusters lies in
the foreground at slightly lower redshifts ($0.04<z<0.03$), while just
a few clusters lie randomly dispersed in the background.

\begin{figure}
\centering
\resizebox{\hsize}{!}{\includegraphics{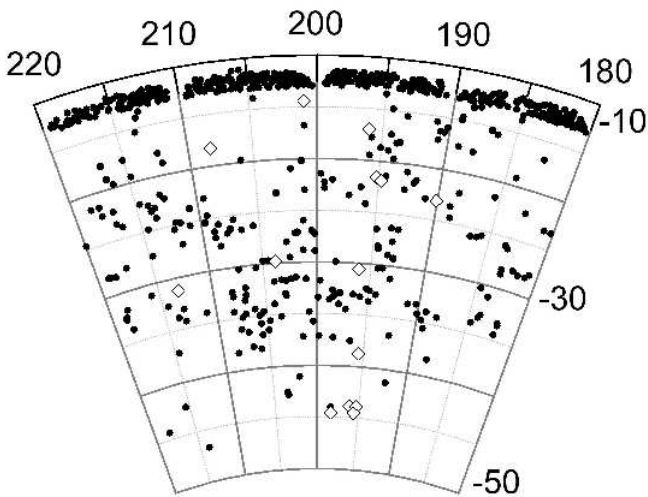}}
\caption{Positions of all known clusters in the area. Empty squares indicate the new detections from this work while filled circles give the positions of all other clusters.}
\label{fig:cl_pos_tot}
\end{figure}

\begin{figure*}[H]
\centering
\includegraphics[angle=270,width=1.0\textwidth]{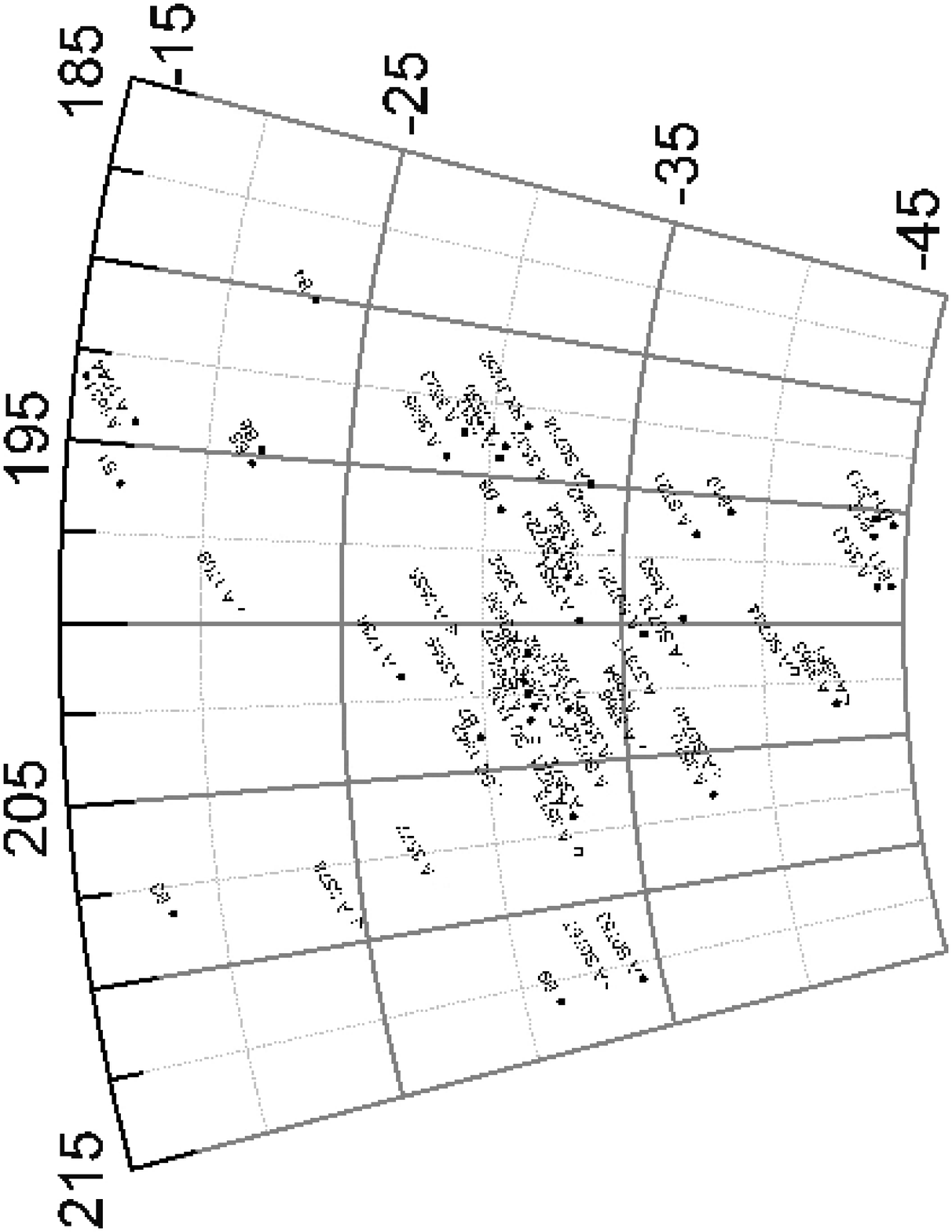}
\includegraphics[width=0.5\textwidth]{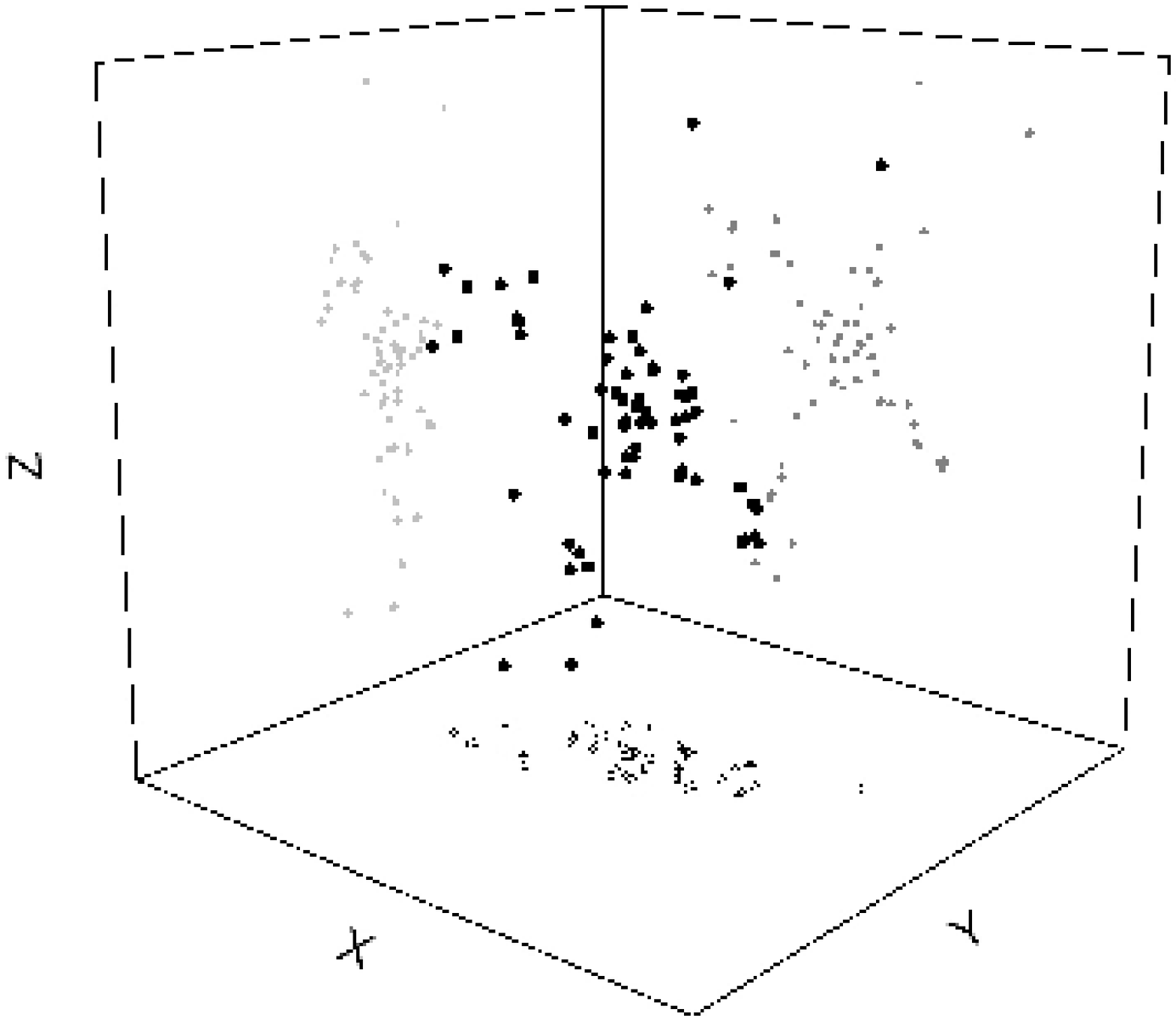}
\caption{Positions of all known clusters in the area inside the
redshift range $0.03\leq z \leq 0.07$. {\bf Top}: RA vs Dec;
filled circles indicate clusters selected and detected in this work,
empty squares represent the clusters selected during the second step
analysis while empty triangles give the positions of all the optically
known clusters, which have not been detected in this survey. {\bf
Bottom}: three-dimensional orthogonal projection; redshift is along
the z axis.}
\label{fig:cl_pos_mio}
\end{figure*}

Figure~\ref{fig:cl_pos_tot} shows the positions of all known clusters
in the area. The high cluster density strip is caused by the Las
Campanas Distant Cluster Survey~\citep{Gon01} which scanned a strip
along the Right Ascension $85 \degr$ long and with a narrow width in
declination ($1.5 \degr$). Figure~\ref{fig:cl_pos_mio} shows the
positions of all known clusters inside the selected redshift range. In
the top panel is a plot of their RA versus their Dec In the
bottom panel the clusters are plotted in a three-dimensional
``orthogonal projection''.

Inside the selected redshift range we detect a total of $54$
extended sources, spread over a total volume of $\approx 1.42 \times
10^6\ {\rm Mpc}^3$. Their distribution is centered at
RA$=201.14\degr$, Dec$=-32.43\degr$ and has a mean redshift of
$z=0.048$.  The projection of the cluster positions on the three
orthogonal planes in the bottom panel of
Fig.~\ref{fig:cl_pos_mio} shows that the superstructure is overall
slightly flattened and elongated; at the center of the super-structure
we can see a dense core, while clusters located outside this core are
positioned along four perpendicular filaments (see projection on plane
y=0), in a strikingly similar way to what seen in simulations of Large
Scale Structure~\citep{Jen98,Yos01}. The SC core seems therefore to be
created at the intersection of almost perpendicular filaments and it
is probably still accreting further clusters through this surrounding
filamentary large scale structure. It is further interesting to notice
that also all our new candidate clusters either lie in the SC core or
along the surrounding filaments. The central dense core of the SC is
located between: $193.14\leq {\rm RA}\leq 208.74$, $-37.64 \leq {\rm
Dec}\leq -27.15$ and $0.0446\leq z\leq 0.0554$. In it we detect (in
a volume of $7.43\times 10^{4}\ {\rm Mpc}^{3}$) $25$ optically known
clusters. This leads to a cluster number density of $\approx 3.4
\times 10^{-4}\ {\rm Mpc}^{-3}$ representing a number density of two
orders of magnitude higher than the mean density of rich Abell
clusters ($\rho_{\rm Abell}\approx 3.4\times 10^{-6}\ {\rm Mpc}^{-3}$)
observed at similar Galactic latitudes~\citep{Abe89}. Our estimate is
about one order of magnitude higher than previous optical measures of
the cluster number density of the Shapley ($\rho_{\rm Shapley}\approx
3.4 \times 10^{-5}\ {\rm Mpc}^{-3}$, ~\citep{Sca89}). If our new
cluster candidates are also included, the cluster number density will
even further increase.

\section{The X-ray luminosity function}
\label{sec:XLF}
A direct measure of the cluster abundance and population in the
Universe is the X-ray luminosity function (XLF), which is defined as
the volume density of clusters per luminosity interval. Comparisons of
the observed cluster abundance distribution as a function of their
luminosity and/or mass with specific cosmological models are
fundamental tests for theoretical models and statistics of large scale
structure. Recently the availability of large cluster samples has made
it possible to reduce the statistical scatter and tightly constrain
the XLF~\citep{Ebe00,Boh01,Mul04}. The strong correlation between the
X-ray luminosity of a cluster and its total mass, makes it then
possible to transform the XLF into a mass function also for large
samples of clusters.

\begin{figure}
\centering
\includegraphics[width=0.5\textwidth]{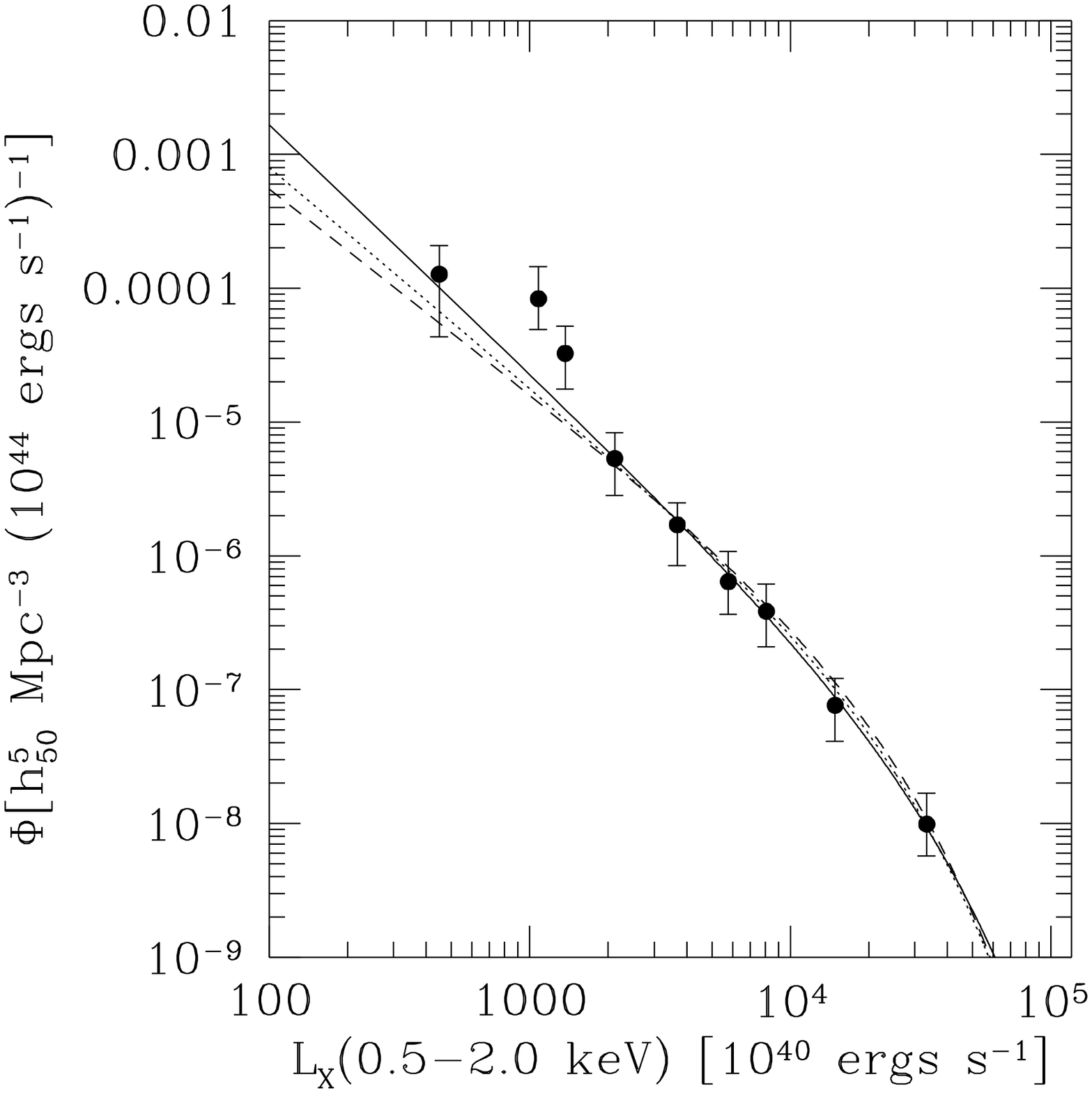}
\includegraphics[width=0.5\textwidth]{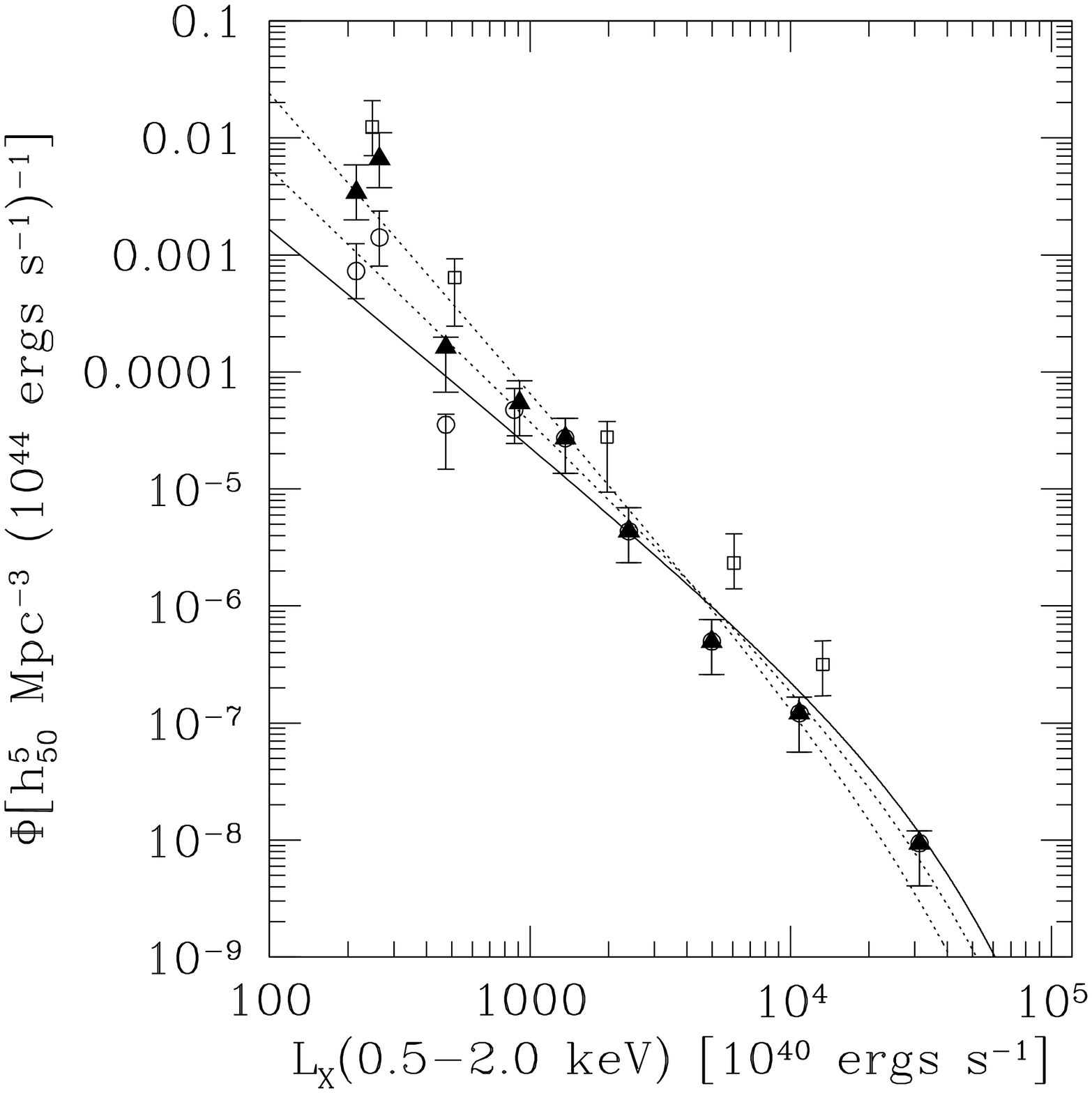}
\caption{X-ray luminosity function. {\bf Top}: Black points represent the binned XLF of our survey. The solid, dotted and dashed lines are the XLFs computed for the BCS, REFLEX and RASS1BS surveys. {\bf Bottom}: ``Lower'' and ``upper'' limit (see text) for the XLF of all the clusters detected in the Shapley redshift range (empty circles and filled triangles, respectively); the two dotted lines are the two best-fit results. The XLF for the central core of the SC is plotted as empty squares. The solid black line is as defined in the top panel.}
\label{fig:XLF}
\end{figure}

We compute the XLF both for all clusters detected in our X-ray survey and for the restricted sample of detected clusters within the Shapley redshift range. The cluster differential luminosity function is defined as:
\begin{equation}
\Phi \left( L_{\rm X}, z\right) = \frac{d^2 N}{dV dL_{\rm X}}\left( L_{\rm X}, z\right)
\end{equation}
where N is the number of clusters of luminosity $L_{\rm X}$ at a redshift $z$ and within a volume $V$. For our sample we compute a binned (with $5$ clusters per bin) luminosity function:
\begin{equation}
\Phi \left( L_{\rm X}, z\right) = \frac{1}{\Delta L} \sum_{i=1}^{N}\frac{1}{V_{\rm max}\left(L_{{\rm X,i}} \right)}
\end{equation}
where the sum over $N$ is the sum over all the clusters falling into
the luminosity interval of the bin and $V_{\rm max}$ is the co-moving
volume in which a cluster of luminosity $L_{{\rm X},i}$ could have
been detected above the flux limit of our survey. Details on how to
compute the luminosity functions are given by~\cite{Boh02}. The top
panel of Fig.~\ref{fig:XLF} shows the resulting binned XLF for all
clusters detected in our survey in the $0.5-2.0\ {\rm keV}$ energy
range ({\it XLF-All}, black points). To allow a direct comparison with
previous surveys the XLFs are computed for an Einstein-de Sitter
cosmology ($\Omega_{\rm M}=1$, $\Omega_{\Lambda}=0$) with $H_0 = 50\
{\rm km\ s^{-1}\ Mpc^{-1}}$. The XLFs estimated from the Brightest
Cluster Sample (BCS+eBCS)~\citep{Ebe98,Ebe00}, REFLEX~\citep{Boh01}
and RASS1BS~\citep{DeG99} surveys are plotted as solid, dotted and
dashed lines, respectively. All three XLFs from the literature provide
an acceptable fit to our data, for clusters with luminosities higher
than $2\times 10^{43}\ {\rm ergs\ s}^{-1}$; a slight excess of low
luminosity clusters is observed in our survey.\\
In the bottom panel we construct a tentative LF from our survey for the Shapley SC alone ({\it XLF-Shapley}). To this aim we use a restricted sub-sample from our survey, given by exclusively clusters which lie in the Shapley redshift range ($0.03\leq z\leq 0.07$). We also add the clusters detected in the {\it second step} analysis. These two samples are obtained with different flux limits. We hence compute two {\it XLFs-Shapley} using two different flux limits: the luminosity of the faintest cluster detected in our analysis  and the flux limit of our survey. The respective XLFs (plotted with empty circles and filled triangles) can therefore be regarded as "upper" and "lower limit" for the {\it XLF-Shapley}. The two dotted lines represent their best-fits. They are both much steeper than values in literature: $\alpha_{\rm L}=2.4, \Phi_{\rm L}^*=3.0\times10^{-8}\ (h_{50}^3\ {\rm Mpc}^{-3}),L_{\rm L}^*=5.7\times10^{44}\ (h_{50}^{-2}\ {\rm ergs\ s}^{-1})$ and $\alpha_{\rm U}=2.15, \Phi_{\rm U}^*=3.7\times10^{-8}\ (h_{50}^3\ {\rm Mpc}^{-3}),L_{\rm U}^*=5.7\times10^{44}\ (h_{50}^{-2}\ {\rm ergs\ s}^{-1})$. 
Even if this restricted sub-sample can definitely not be
considered complete, and therefore no quantitative result can be drawn
from it, we use it to draw generic conclusions on the luminosity
population of the clusters in the Shapley SC.
In \S~\ref{sec:morph_anal} we measured an over-abundance of
clusters in the Shapley region, and especially in the core, with respect 
to the average number
density of Abell clusters. We can now say that while the bright
cluster population of the whole Shapley SC is consistent with what is observed in
lower density environments, low luminosity clusters are over-abundant;
low luminosity clusters are hence the main cause of the observed
excess in the cluster number density of this super-structure. \\
If we restrict our analysis to the central dense core of the SC alone 
($0.0446\leq z\leq 0.0554$, $193.14\leq {\rm RA}\leq 208.74$ and $-37.64 \leq 
{\rm Dec}\leq -27.15$, plotted as empty squares in the bottom panel of 
Fig.~\ref{fig:XLF}), while we still observe an over-abundance of low-luminosity 
objects, high luminosity clusters also are in excess with respect to the background distribution.
Since we expect strongly interacting and merging clusters to be X-ray luminous 
objects, because the X-ray luminosity is enhanced during a merger, 
the external regions of the SC are most probably still in a process of accreting 
low luminosity, small objects through the surrounding filaments.
Major close encounters, at this point of the SC formation, affect 
the luminosity of the cluster population only in the central densest core.

\subsection{Optical versus X-ray cluster properties}
\label{sec:opt_vs_X}
From~\cite{Gir00,Gir98} we extracted the values of the optical luminosities in the B$_{\rm j}$ band computed within the cluster virial radius, together with the values of the virial radii available for the clusters in our sample.

\begin{figure}
\centering
\resizebox{\hsize}{!}{\includegraphics{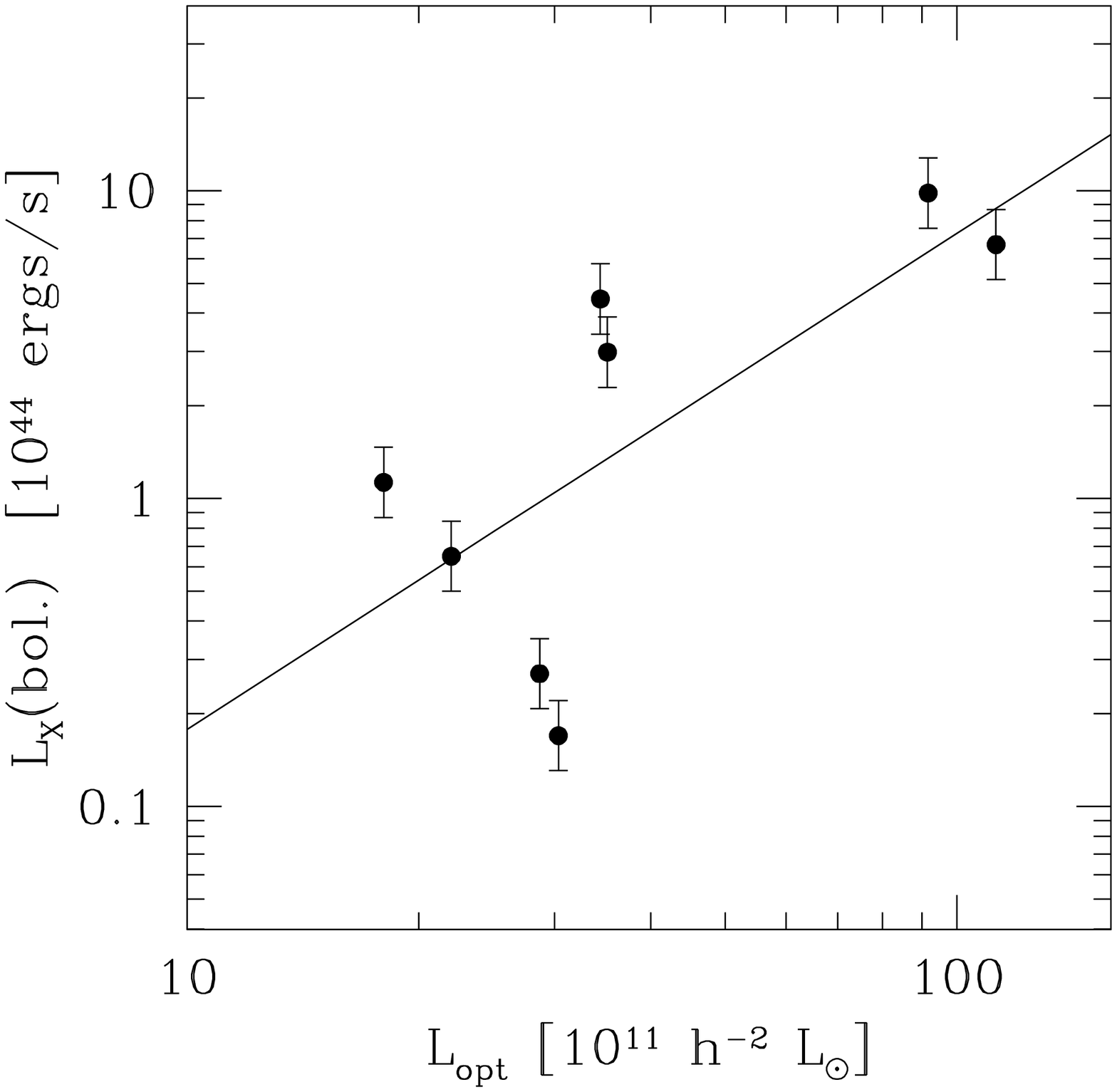}\includegraphics{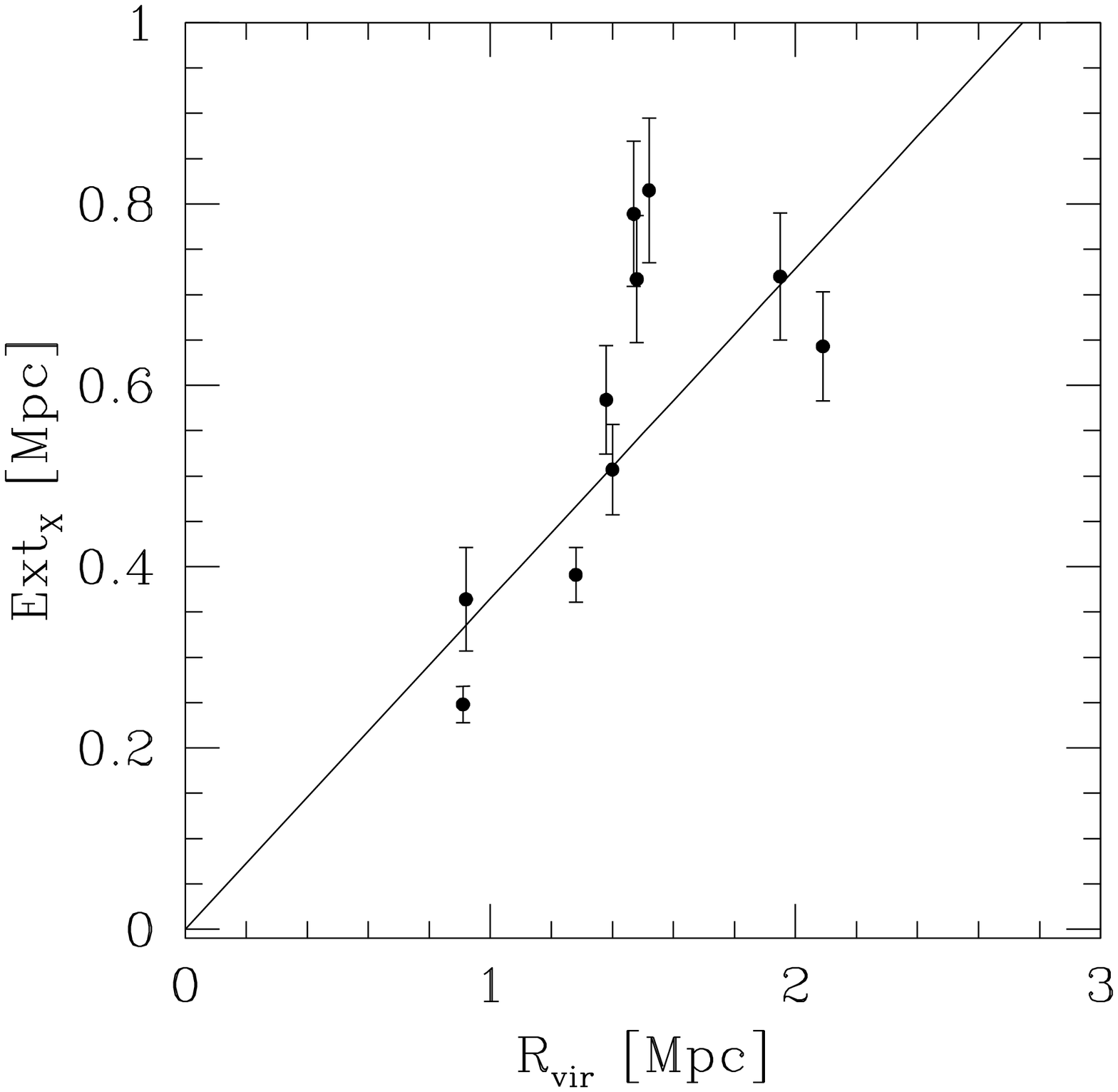}}
\caption{{\bf Left panel}: Bolometric X-ray luminosity versus optical luminosity. {\bf Right panel}: Extension of the cluster from our X-ray detection versus virial radius. Best-fit estimates are plotted as solid lines.}
\label{fig:LL-RR}
\end{figure}

In the left panel of Fig.~\ref{fig:LL-RR} we plot the bolometric X-ray luminosity of the clusters versus their optical B$_{\rm j}$ luminosity. The solid line represents the result of the fit: 
$${\rm L}_{\rm X}= 10^{\left(-2.36\pm1.09\right)}{\rm L}_{\rm opt}^{\left(1.61\pm 0.68\right)}$$
In the right panel is the extent of the clusters as computed from our detection algorithm against their virial radius, both in units of Mpc. The best-fit line is plotted as a solid line.
$${\rm Ext}_{\rm X}=\left(0.36\pm 0.14\right)\cdot R_{\rm vir}$$
For both relations there are signs of a trend, but the number of objects in our sample for which optical data are available is unfortunately too low to be able to give any definite result. Similar, more accurate trends for the ${\rm L}_{\rm X}-{\rm L}_{\rm opt}$ relation, based on a much larger sample, were recently found by~\cite{Pop05}.

\subsection{Merger rates/cosmology}
\label{sec:merger}
The dynamical state of clusters of galaxies depends on the matter
density $\Omega_m$. This was first shown in an analytical approach by~\cite{Ric92} and modified later by other groups taking into account
time variations for substructure to be erased ~\citep{Kau93,Lac93,Nak95}. 
As the amount of substructure is expected to increase with $\Omega_m$, 
several X-ray
samples have been used to determine the frequency of substructure (i.e.
the merger rate): \cite{Moh95} 50-70\%;
\cite{Jon99} 40\%;
\cite{Sch01} $52\pm7$\%. The results show some scatter in
the numbers because different methods have been used and the rates
also depend on the resolution of the instrument as well as on several 
other parameters.

The merger rates, however, do not only depend on the overall mass
density but they are expected to depend also on the local density. 

To determine the merger rate for the Shapley super-cluster we classified
the clusters according to their morphology. $25$ clusters have sufficient photons to determine the morphology; to these we added the four new cluster detection which were confirmed clusters in the optical (B1, B4, B5 and B9).  
In the last column of Table~\ref{tab:ris_fin} these clusters are flagged with an ``M'' if they show any signature of mergers (i.e. from clearly interacting structures, to strongly irregular, filamentary and unsymmetrical and deformed morphologies).
We find that out of the 29 clusters 22 clusters show
such signatures, i.e. a fraction of 76\% of the clusters are not
relaxed. 
We are aware that
such a classification is just a somewhat arbitrary estimate, but we believe it to be sufficient for our actual purpose, which is to compare it with 
several previous estimates, which were all derived with different methods. 

Obviously, the merger rate in the
Shapley super-cluster is higher than in all the general cluster samples
by~\cite{Moh95,Jon99,Sch01}. This
confirms the expectations of an increasing merger rate with
increasing local density, which also has been found by~\cite{Sch01}. 
This dependence provides a possibility to determine the
total mass of the Shapley super-cluster in an independent way. 
From simulations the merger rate could be determined at different local
mass densities. With the same method the merger rate could be
determined from an observed general cluster sample and from our Shapley
sample. The comparison would yield the total mass density and hence
also the total mass of the Shapley super-cluster. We hope that such a
set of simulations will be available soon.

\section{The cumulative mass profile}
\label{sec:mass}
\begin{figure}
\centering
\resizebox{\hsize}{!}{\includegraphics{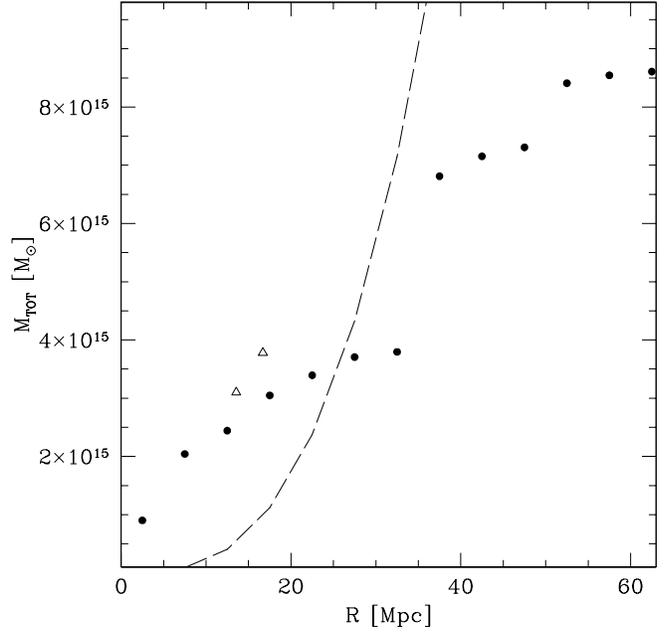}}
\caption{Cumulative mass profile of detected clusters in the Shapley redshift range (black dots). The dashed line represents the expected total mass in a homogeneous Universe of density $\rho_c\Omega_0$. The empty triangles are the values of the virial mass computed considering the single clusters as test particles. See Table~\ref{tab:masse} for numerical values at different radii.}
\label{fig:massa_vs_r}
\end{figure}

We then compute the cumulative mass profile of all detected clusters within the restricted Shapley redshift range ($0.0446\leq z\leq 0.0554$); this is defined as the measured (from X-rays) matter distribution in the Shapley SC as a function of the co-moving 3-D distance from A~3558. The total mass for each cluster is estimated from their X-ray luminosity using the $L_{\rm X}/M$ relation by~\cite{Rei02}:
\begin{equation}
{\rm Log}\left[ \frac{L_{\rm X} \left( 0.1-2.4\ {\rm keV}\right)}{h^{-2}_{50}\ {\rm ergs\ s}^{-1}}\right]=A+\alpha\ {\rm Log}\left( \frac{M_{200}}{h^{-1}_{50}\ M_{\odot}}\right)
\label{eq:total_mass}
\end{equation}
with $\alpha=1.807\pm0.084$ and $A=-22.053\pm1.251$. 
The cumulative mass profile is then estimated by summing the masses 
of all clusters detected within the selected Shapley redshift range 
(including clusters detected in the second step analysis).
In Fig.~\ref{fig:massa_vs_r} we compare the resulting mass profile, plotted 
as black dots, with the expected virial and cosmic masses, in a similar
way as done by~\cite{Ett97}.
The dashed line represents the expected mass, within the same distance from A~3558, in a homogeneous Universe of density $\rho_c\Omega_m$:
\begin{equation}
M_0=\frac{H_0^2}{2G}R^3\Omega_0 (1+z)^3
\end{equation}
The empty triangles represent the values of the virial mass computed considering the single clusters as test particles using:
\begin{equation}
M_{\rm Vir}=\frac{3\pi}{G}\frac{\sigma^2}{\langle 1/R_{\perp i,j} \rangle}
\end{equation}
where $\sigma$ is the standard deviation of the distribution of the
observed velocities of the clusters and $\langle 1/R_{\perp i,j} \rangle$ is the
harmonic mean radius. Table~\ref{tab:masse} lists the values of the
total X-ray mass ($M_{\rm X}$), of the mass expected within a
homogeneous Universe ($M_0$) and of the virial mass ($M_{\rm Vir}$),
computed within four increasing radii. While we do measure an
over-abundance of X-ray mass up to ${\rm R}=25\ {\rm Mpc}$ with
respect to $M_0$, the gravitating mass is still slightly lower than
the virial value. This does not come as a surprise since the SC has not
yet reached a virial equilibrium~\citep{Ett97,Bar00} and seems to be still 
accreting matter through the surrounding filamentary large scale 
structure (see \S~\ref{sec:morph_anal}).

Our X-ray cumulative mass profile is strictly a lower limit 
since it is computed considering only the sum of the gas and 
gravitating mass of each cluster. We neglect any contribution from a 
possible intra-supercluster-medium and dark matter and do not consider 
matter in the outer regions of clusters, where the X-ray emission detected in the RASS 
survey gets confused with the background. Furthermore, some of the known optical clusters 
are still undetected in X-rays; this could be due to the fact that these clusters are chance 
alignments of galaxies along the line of sight, or that they represent extremely faint structures, 
below our flux limit. 

As seen in \S~\ref{sec:XLF}, the Shapley SC shows an extremely rich population of faint extended sources; 
this population provides a substantial portion of the mass content of the whole SC. We hence stress the 
necessity to explore this region, as well as other similarly crowded ones, to increasingly lower X-ray 
luminosities in order to obtain an exhaustive knowledge of their mass and of their cluster population. 

We estimate the total mass of the SC using all 
clusters detected in the Shapley redshift range.  To this aim we use the 
{\it XLF-Shapley} measured in \S~\ref{sec:XLF} to compute the total $L_{\rm X}$ 
emitted by the SC. We then estimate the total SC X-ray mass using 
eq.~\ref{eq:total_mass}. If we include clusters as faint as those detected in our second step analysis ($L_X\approx 2\times 10^{42}\ {\rm ergs\ s}^{-1}$) the resulting SC total mass range is $M_{\rm X}=5-8\times 10^{15}\ M_\odot$ (from the lower and upper limit of the {\it XLF-Shapley}). Integrating to even lower luminosities, including structures as faint as $L_X\approx 1\times 10^{42}\ {\rm ergs\ s}^{-1}$, would further double the value of the total SC mass.

\begin{table}
\caption{Total Mass Estimates}
\label{tab:masse}      
\centering
\begin{tabular}{ccccc}
\hline\hline       
R	&N$_{\rm cl}$ & $M_{\rm X}$	& $M_0$	& $M_{\rm Vir}$ \\ 
(Mpc)	& & $\left(10^{14}\ M_\odot\right)$	& $\left(10^{14}\ M_\odot\right)$	& $\left(10^{14}\ M_\odot\right)$ \\
\hline  
$13.6$ &$7$	&$23.5$  &$5.2$     &$31.0$ \\
$16.7$ &$9$	&$28.8$  &$9.7$     &$37.8$ \\
$40.0$ &$28$	&$57.7$  &$133.2$   &$173.6$ \\
$60.8$ &$31$	&$72.3$  &$467.9$   &$180.9$ \\
\hline                  
\end{tabular}
\begin{list}{}{}
\item Col. (1): Co-moving distance from A~3558 within which $M_{\rm X}$, $M_0$ and $M_{\rm vir}$ (Cols. (3) - (5), respectively) are estimated. $M_{\rm X}$ is computed as the sum of the masses of all clusters detected in the Shapley redshift range (including clusters detected in the second step analysis). In Col. (2) is the number of clusters analyzed within each radius.
\end{list}
\end{table}

\section{Results and discussion}
\label{sec:results}
We have surveyed the X-ray emission of a large area surrounding the
Shapley SC using RASS-III data. To this purpose we used a detection 
algorithm, explicitly written to detect 
nearby extended, interacting and/or irregular sources.  
Within the analyzed region our algorithm initially detected $56\%$ of all optically
known clusters below redshifts of $0.07$. For $9$ of them this was the first
X-ray detection. Only $3$ clusters with previous X-ray detection were not
found in our survey.
For each optically known cluster which was not detected, we performed a more
accurate second step analysis lowering our flux limit. In this way we could
detect X-ray emission from $40$ SC members already known in the optical ($\sim
80\%$ of all known extended structures); for half of these clusters this is
the first detection in X-rays. We also isolated $14$ new cluster
candidates. Follow-up observations have shown at least $4$ of these candidates
have optical counterparts at their X-ray positions.

The analysis of such a wide area has allowed us to study the cluster distribution not only in the SC core but also in its surroundings. 
The overall structure of the SC looks slightly flattened and elongated;
showing an extremely dense core at the center ($193.14\leq {\rm
RA}\leq 208.74$, $-37.64 \leq {\rm Dec}\leq -27.15$ and
$0.0446\leq z\leq 0.0554$). The clusters surrounding the dense core 
lie along perpendicular filaments similarly to what is
shown by large scale structure simulations; they are most probably moving 
towards the SC core, attracted by its gravitational potential.

Compared to previous X-ray analysis of the region~\citep{Ray91,Ett97}, 
we have largely increased the number of clusters in the area detected 
in X-rays (thanks to both the wide area analyzed and the low flux limit 
reached in our second step analysis). This has led to an estimate of the local 
cluster number density  more than one order of magnitude higher 
than previous optical estimates~\citep{Sca89} both in the core region 
and in the much less dens surroundings.

In our measure of the X-ray mass we kept a strictly conservative approach, 
since we neglected any contribution
from baryonic and dark matter between the clusters and from the outer
regions of the clusters. 
In the SC core we measure an over-abundance of X-ray mass with respect to 
what expected
in a homogeneous Universe, while the mass in clusters is slightly
lower than what is needed to account for the virial value, sign that the 
SC, even in its central core, is not yet virialized. \\
A direct comparison with previous works is not straightforward because of 
several fundamental differences which differentiate each work 
(i.e. number of clusters, sky region, cosmology, etc.). 
As a general statement we can though say that our X-ray mass estimate 
is consistent with that measured by~\cite{Ett97} within a distance of 
$20\ {\rm Mpc}$ from A~3558, and higher only up to a factor of $1.6$ going 
out to larger radii. Our estimates is though based on a higher number 
of clusters; 
at the same time~\cite{Ett97} extrapolate the cluster mass 
out to much larger radii. Both X-ray values barely reach
the lower limits of the gravitating mass estimated by~\cite{Ray91,Qui95}
and~\cite{Bar00}
using different approaches (excess in the number of galaxies, 
virial estimate of each cluster mass, etc.).
Once deep X-ray observations of a larger number of Shapley clusters will 
be available, we will be able to obtain more accurate mass measures, out
to increasingly larger radii, even for the faintest clusters in the region.

Rich, X-ray luminous clusters are common objects in the 
central dense SC core~\citep{Ray91}. The analysis of a wide sky area 
performed in this work has highlighted 
that these are instead much more rare objects in the SC 
outskirts. Faint clusters and poor groups are instead over-abundant
both in the central and in the outer regions of the Shapley cluster region compared 
to what is measured in less dense environments. 
Such objects are hence expected to provide an extremely important
addition to the SC mass content. To date we have examined only
the luminous tail of the X-ray luminosity function for the Shapley
region. We therefore underline the importance of exploring these
crowded regions to fainter luminosities, which would significantly
increase our knowledge of their mass and population content.

\begin{acknowledgements}
This paper is based on observations obtained from the {\it ROSAT} Public Data Archive and on observations 
made with ESO Telescopes at the La Silla Observatory.
We thank MPE and ESO for maintaining their archives active and running 
and the NASA HPCC ESS group at the University of Washington for making 
publicly available their software tools. 
This work has been supported by NASA grants NAS8-39073, by FWF grant P15868 of the Austrian Science Foundation and
by the Deutsche Forschungsgemeinschaft (DFG) under the project ER 327/2-1.\\
We thank the referee Stefano Ettori for his help and useful comments.
\end{acknowledgements}

\bibliographystyle{aa}
\bibliography{biblio_sha}

\begin{sidewaystable*}
\vspace{0.5truecm}
\begin{minipage}[t][180mm]{\textwidth}
\begin{scriptsize}
\caption{Detected Sources}             
\label{tab:ris_fin}
\centering
\renewcommand{\footnoterule}{} 
\begin{tabular}{lcccrrrrrrrrr}
\hline\hline       
Cluster & RA; Dec & z &$n_{\rm H}$& Radius & HR & Count Rate & $f_{\rm X}$ & $f_{\rm bol}$ & $L_{\rm bol}$ & $kT$ & Refer. & Merger\\
\cline{8-9}  & (J2000.0) & &$(10^{20}$ &  &  &  & \multicolumn{2}{c}{$(10^{-12}$} & $(10^{44}$ & \\
 & (deg) &  &${\rm cm}^{-2})$ & (arcmin) &  & $\left( {\rm counts\ s}^{-1}\right)$ & \multicolumn{2}{c}{${\rm ergs\ cm}^{-2}\ {\rm s}^{-1})$} & ${\rm ergs\ s}^{-1})$ & (keV)& &\\ 
\hline  
RX J1252.5-3116 &$193.143;-31.266$  &$0.0535$   &$5.55$  &$5.8$  &$+0.80$  &$0.69\pm0.08$ &$16.09$ &$30.33 $ &$1.95 $ &$3.8$ &  & \\
Abell 1631      &$193.242;-15.379$  &$0.0462$   &$3.94$  &$9.4$  &$-0.24$  &$0.15\pm0.01$ &$3.43 $ &$5.65  $ &$0.27 $ &$2.8$ & A & \\
Abell 3528      &$193.640;-29.126$  &$0.0528$   &$6.07$ &$12.24$ &$+0.50$  &$1.04\pm0.11$ &$24.32$ &$47.02 $ &$2.95 $ &$4.0$ & B & M \\
Abell 3528N     &$193.598;-29.010$  &$0.0528$   &$6.07$  &$6.01$ &$+0.70$  &$0.44\pm0.04$ &$10.53$ &$18.07 $ &$1.13 $ &$3.4$ & B & \\
Abell 3528S     &$193.673;-29.231$  &$0.0528$   &$6.07$  &$6.33$ &$+0.68$  &$0.51\pm0.05$ &$12.20$ &$20.97 $ &$1.31 $ &$3.1$ & B & \\
Abell 3530      &$193.917;-30.367$  &$0.0537$   &$5.75$  &$9.7$  &$+0.42$  &$0.39\pm0.03$ &$9.24 $ &$16.10 $ &$1.04 $ &$3.2$ & C & M \\
Abell 1644      &$194.332;-17.381$  &$0.0473$   &$4.05$  &$14.8$ &$+0.49$  &$1.91\pm0.16$ &$41.28$ &$89.21 $ &$4.45 $ &$5.0$ & D & M \\
Abell 3532      &$194.336;-30.375$  &$0.0554$   &$5.84$  &$11.4$ &$+0.45$  &$0.93\pm0.07$ &$21.35$ &$43.18 $ &$2.99 $ &$4.4$ & A & M \\
Abell 3535      &$194.446;-28.536$  &$0.0652$   &$6.49$  &$5.4$  &$+0.11$  &$0.090\pm0.006$&$2.14 $ &$2.99  $ &$0.21 $ &$1.8$ & & M \\
Abell S0718     &$194.963;-33.661$  &$0.0478$   &$5.57$  &$3.7$  &$+0.56$  &$0.060\pm0.006$&$1.43 $ &$1.89  $ &$0.097$ &$1.1$ && \\
Abell S0721     &$196.513;-37.642$  &$0.0490$   &$5.13$  &$10.4$ &$+0.09$  &$0.33\pm0.03$ &$7.76 $ &$12.10 $ &$0.65 $ &$2.5$ && \\
Abell S0724     &$198.286;-32.994$  &$0.0493$   &$4.86$  &$5.0$  &$-0.90$  &$0.090\pm0.007$&$2.05 $ &$2.75  $ &$0.15 $ &$1.3$ && \\
Abell 3548      &$198.379;-44.063$  &$0.0000$   &$8.43$  &$6.5$  &$+0.51$  &$0.17\pm0.02 $&$4.26 $ &$6.08  $ &$0.34 $ &$1.9$ && \\
Abell 3553      &$199.805;-37.179$  &$0.0487$   &$4.73$  &$3.5$  &$+0.33$  &$0.058\pm0.003$&$1.35 $ &$1.79  $ &$0.095$ &$1.1$ && \\
Abell 3554      &$199.865;-33.497$  &$0.0470$   &$4.7 $  &$8.5$  &$-0.17$  &$0.128\pm0.016$&$2.91 $ &$3.95  $ &$0.20 $ &$1.5$ && \\ 
Abell S0729     &$200.358;-35.816$  &$0.0499$   &$4.15$  &$2.8$  &$+0.39$  &$0.055\pm0.008$&$1.27 $ &$1.67  $ &$0.093$ &$1.1$ && \\
Abell 3556      &$201.001;-31.656$  &$0.0479$   &$4.07$  &$7.1$  &$-0.20$  &$0.077\pm0.003$&$1.72$  &$3.26  $ &$0.17 $ &$3.8$ & E& M \\
Abell 1736      &$201.758;-27.153$  &$0.0453$   &$5.38$  &$12.2$ &$+0.38$  &$1.18\pm0.20$ &$27.54$ &$50.17 $ &$2.27 $ &$3.5$ & D& \\
Abell 3558      &$202.011;-31.493$  &$0.0480$   &$3.89$  &$13.1$ &$+0.60$  &$2.72\pm0.11$ &$57.87$ &$131.0 $ &$6.68 $ &$5.5$ & D& M \\
SC 1327-312     &$202.514;-31.664$  &$0.0495$   &$3.89$  &$11.4$ &$+0.19$  &$0.55\pm0.07$ &$12.25$ &$23.16 $ &$1.27 $ &$3.8$ & E& M \\
SC 1329-314     &$202.875;-31.812$  &$0.0446$   &$3.87$  &$7.2$  &$+0.25$  &$0.265\pm0.026$&$5.84 $ &$11.59 $ &$0.52 $ &$4.2$ & E & M \\
RXJ1332.2-3303  &$203.109;-33.092$  &$0.0470$   &$4.08$  &$9.6$  &$+0.38$  &$0.55\pm0.06$ &$11.90$ &$21.27 $ &$1.06 $ &$3.0$ && \\
Abell 3562      &$203.446;-31.687$  &$0.0490$   &$3.83$  &$14.0$ &$+0.31$  &$1.35\pm0.11$ &$29.16$ &$61.73 $ &$3.31 $ &$4.8$ & D & M \\
Abell 3563      &$203.394;-42.502$  &$0.0000$   &$7.82$  &$5.4$  &$+0.31$  &$0.130\pm0.013$&$3.19 $ &$4.41  $ &$0.25 $ &$1.6$ && \\
Abell 3570      &$206.708;-37.874$  &$0.0366$   &$4.47$  &$6.5$  &$+0.11$  &$0.159\pm0.009$&$3.61 $ &$4.81  $ &$0.13 $ &$1.2$ && \\
Abell 3571      &$206.860;-32.850$  &$0.0391$   &$3.91$  &$14.0$ &$+0.61$  &$5.3\pm0.5  $&$110.9$ &$282.0 $ &$9.82 $ &$6.9$ & D & \\
Abell S0758     &$213.065;-34.314$  &$0.0380$   &$4.78$  &$4.4$  &$+0.13$  &$0.083\pm0.009$&$1.97$  &$2.61  $ &$0.082$ &$1.0$ && \\
B1$^{\mathrm{(a)}}$&$196.080;-17.001$  &$--$   &$5.26$  &$5.6$  &$+0.32$  &$0.319\pm0.012$&$7.53$  &$11.76 $ &$0.66$  &$2.5$ & & M \\
B2$^{\mathrm{(b)}}$&$200.977;-14.437$  &$--$   &$4.56$  &$5.2$  &$+0.40$  &$0.40\pm0.03$ &$9.27$  &$15.07 $ &$0.84$  &$2.7$ && \\
B3$^{\mathrm{(b)}}$&$208.265;-18.299$  &$--$   &$7.67$  &$5.5$  &$+0.33$  &$0.078\pm0.007$&$1.91$  &$2.55  $ &$0.14$  &$1.3$ && \\
B4$^{\mathrm{(a)}}$&$190.088;-23.131$  &$--$   &$6.53$  &$3.2$  &$+0.39$  &$0.075\pm0.007$&$1.79$  &$2.39  $ &$0.13$  &$1.3$ & & M \\
B5$^{\mathrm{(a)}}$&$195.215;-21.608$  &$--$   &$7.59$  &$6.6$  &$+0.15$  &$0.078\pm0.012$&$1.91$  &$2.55  $ &$0.14$  &$1.3$ & & M \\
B6$^{\mathrm{(a)}}$&$194.795;-21.911$  &$--$   &$7.49$  &$5.6$  &$+0.28$  &$0.150\pm0.008$&$3.66$  &$5.14  $ &$0.29$  &$1.8$ && \\
B7$^{\mathrm{(c)}}$&$203.839;-29.862$  &$--$   &$4.02$  &$6.9$  &$+0.13$  &$0.29\pm0.04$ &$6.63$  &$10.12 $ &$0.57$  &$2.3$ && \\
B8$^{\mathrm{(b)}}$&$196.085;-30.593$  &$--$   &$5.73$  &$8.8$  &$+0.38$  &$0.209\pm0.009$&$4.94$  &$7.21  $ &$0.40$  &$2.0$ & & M \\
B9$^{\mathrm{(c)}}$&$213.224;-31.277$  &$--$   &$4.14$  &$5.6$  &$+0.16$  &$0.139\pm0.015$&$3.12$  &$4.30  $ &$0.24$  &$1.6$ && \\
B10$^{\mathrm{(b)}}$&$195.548;-38.760$  &$--$   &$6.89$  &$4.8$  &$+0.11$  &$0.073\pm0.008$&$1.76$  &$2.34  $ &$0.13$  &$1.2$ && \\
B11$^{\mathrm{(c)}}$&$198.339;-44.591$  &$--$   &$6.74$  &$5.6$  &$+0.17$  &$0.09\pm0.01$ &$2.15$  &$2.89  $ &$0.16$  &$1.4$ && \\
B12$^{\mathrm{(c)}}$&$196.136;-43.868$  &$--$   &$8.32$  &$7.0$  &$+0.44$  &$0.22\pm0.02$ &$5.55$  &$8.22  $ &$0.46$  &$2.1$ && \\
B13$^{\mathrm{(c)}}$&$195.338;-43.902$  &$--$   &$7.64$  &$6.4$  &$+0.31$  &$0.087\pm0.006$&$2.13$  &$2.85  $ &$0.16$  &$1.4$ && \\
B14$^{\mathrm{(b)}}$&$195.582;-44.535$  &$--$   &$7.29$  &$4.8$  &$+0.24$  &$0.081\pm0.009$&$1.97$  &$2.63  $ &$0.15$  &$1.3$ && \\
\hline                  
\end{tabular}
\end{scriptsize}
\begin{tiny}
\begin{list}{}{}
\item Col. (1): Source name. The last $14$ sources labeled with B followed by increasing identification numbers are the new X-ray detections. (a) Confirmed cluster. (b) Possible cluster. (c) Non cluster source. Col. (2): RA and Dec (J2000). Col. (3): Redshift. Col. (4); Galactic HI column density in units of $10^{20}\ {\rm cm}^{-2}$~\citep{Dic90}. Col. (5): Source radius (arcmin). Col. (6): Hardness ratio. Col. (7): Source count rate in the $0.5-2.0\ {\rm keV}$ energy band and its error. Cols. (8) - (9): $f_{\rm X}$ in the $0.1-2.4\ {\rm keV}$ energy band and the bolometric flux, both in units of $10^{-12}\ {\rm ergs\ cm}^{-2}\ {\rm s}^{-1}$. Col. (10): Bolometric luminosity in units of $10^{44}\ {\rm ergs\ s}^{-1}$; for the new X-ray detections a redshift z$=0.05$ has been assumed. Col. (11): Cluster average temperature in keV. Col. (12): Literature reference for the cluster temperature: A~\citep{Wu99}, B~\citep{Sch96}, C~\citep{Ett97}, D~\citep{Mar98b}, E~\citep{Han99}. Col. (13): Flag for clusters showing signatures of merger. 
\end{list}
\vfill
\end{tiny}
\end{minipage}
\end{sidewaystable*}

\appendix
\section{Optical properties of B1, B4-B9 and B11-B14}
\label{app:properties}
\begin{figure*}
\centering
\includegraphics[height=6.5truecm]{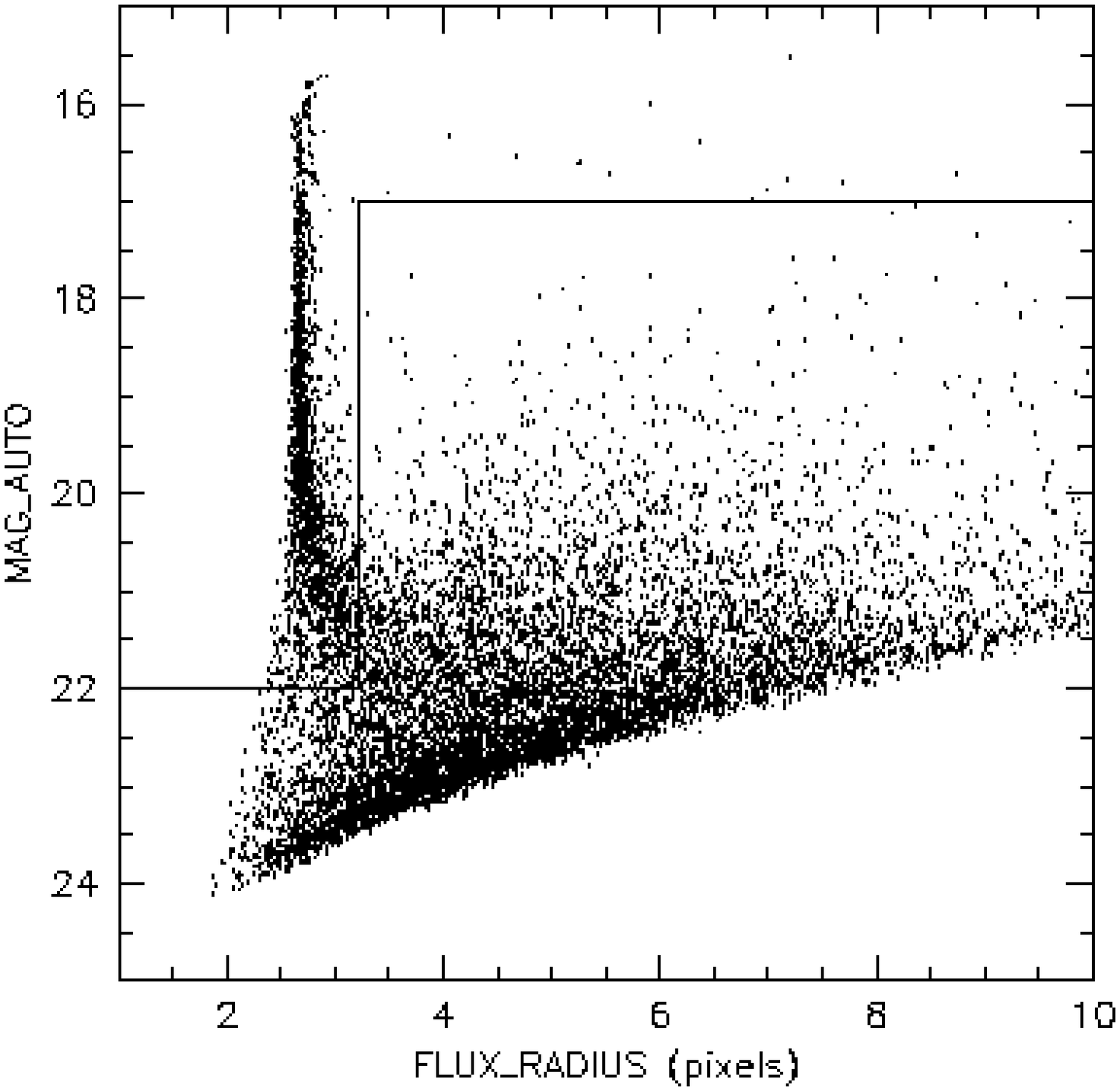}
\includegraphics[height=6.5truecm]{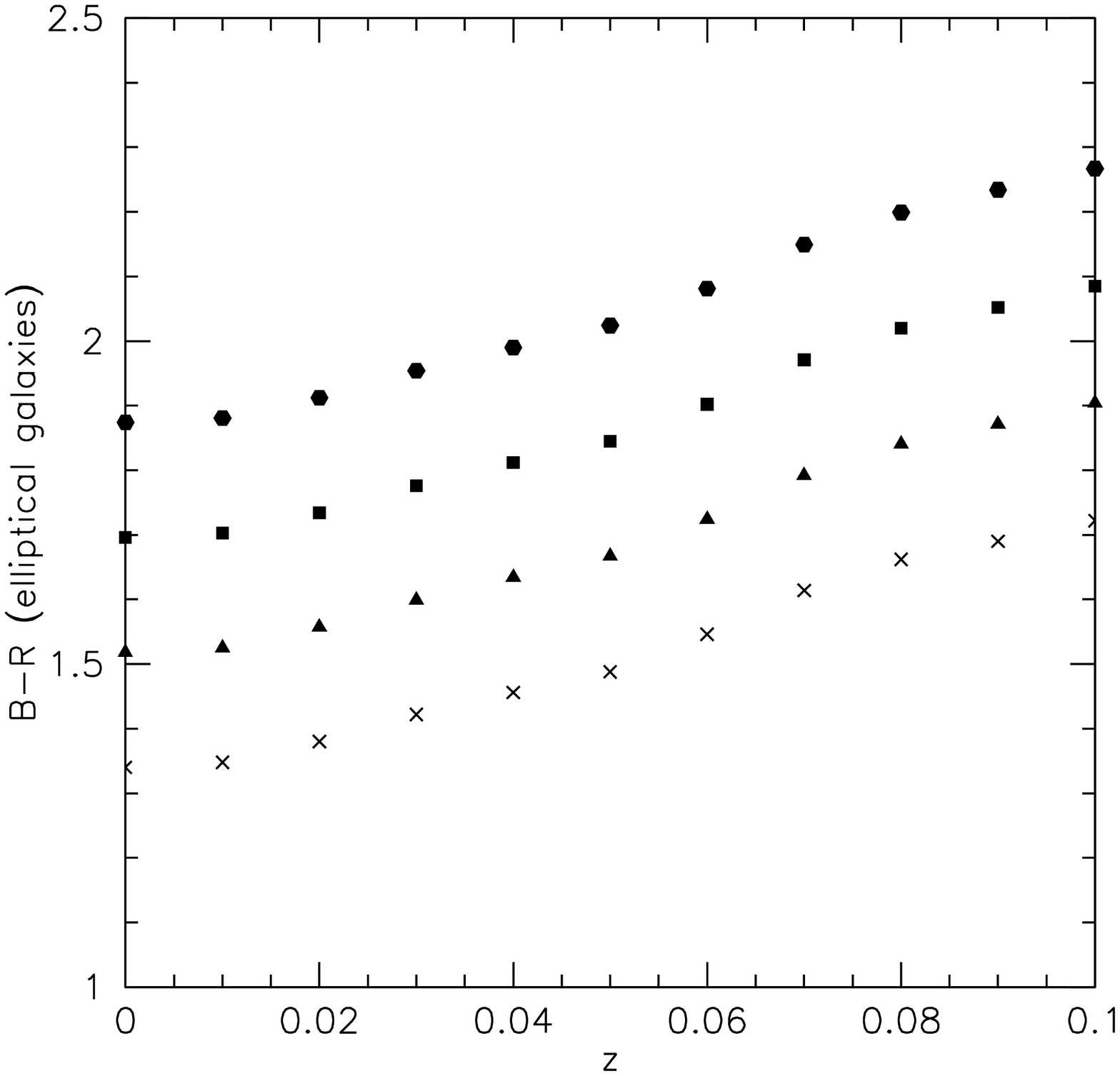}
\caption{Shown in the {\bf left panel} is a half-light radius (SExtractor
FLUX\_RADIUS parameter) vs. $R$ magnitude (SExtractor MAG\_AUTO
parameter) diagram from all detected objects in the field of B5. The
bright stars are located in a narrow locus around $r_{\rm h}\approx
3.2$. In our analysis we reject all objects in the box shown around
the stellar locus and we apply an additional cut for bright galaxies
($R<17$).  The contamination of our galaxy catalog with stars in the
faint part of the stellar locus should not affect our analysis
significantly. We reject bright objects to avoid the domination of our
light maps by individual galaxies. Identical plots were created and examined
for all fields (the cut for bright objects is always at $R=17$). 
The {\bf right panel} shows $B-R$ colors
of elliptical galaxies as a function of redshift.  To create the plot
we used galaxy templates from~\cite{Bru93} with an exponential law
(time-scale 1 Gyr) for the star formation rate.  We applied the
reddening law from~\cite{Cal00} with extinctions $A_v=0.0$ (crosses),
$A_v=0.5$ (triangles), $A_v=1.0$ (squares) and $A_v=1.5$
(hexagons). Given this plot together with our $B$ and $R$ band data
we search for Red Cluster Sequences of
potential SC members in the color range $1.5<B-R<2$.}
\label{fig:rhmagbmr}
\end{figure*}

\begin{figure*}
\centering
\resizebox{\hsize}{!}{\includegraphics[angle=-90,width=0.2\textwidth]{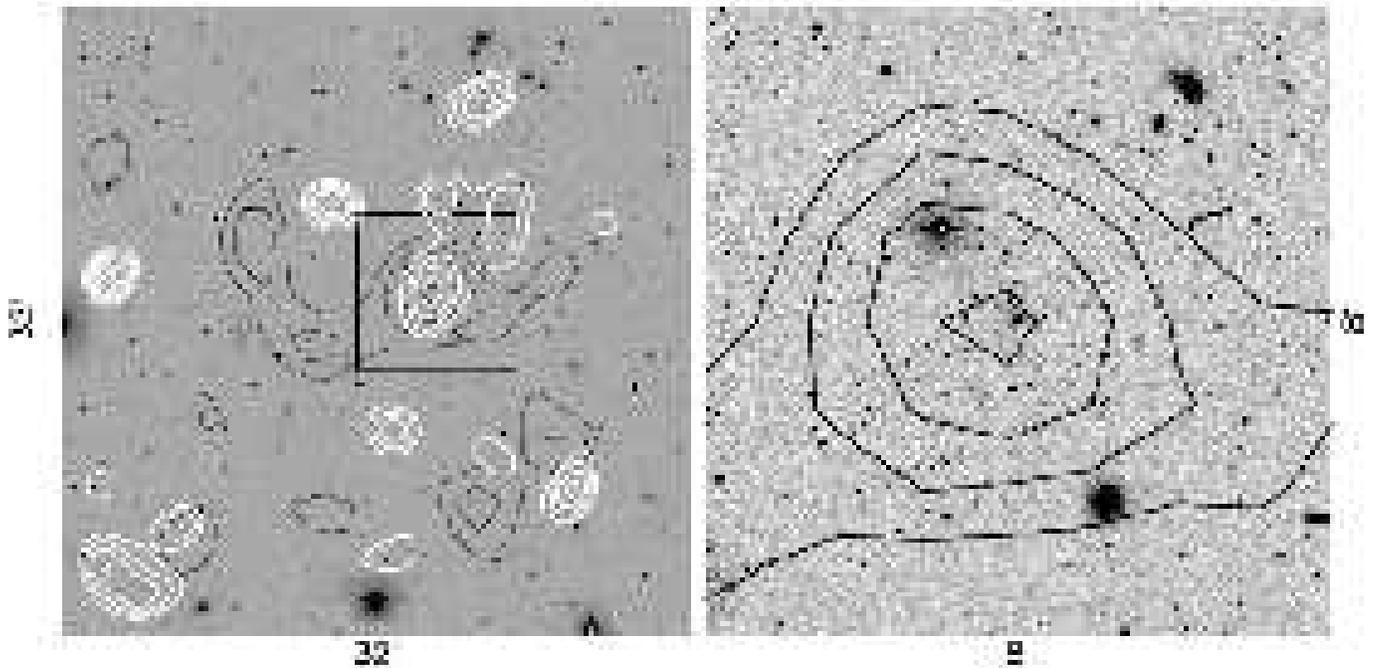}}
\caption{{\bf Left panel}: shown is the full WFI $R$-band field around the X-ray detection B1. RA increases to the left and Dec to the top. Black contours mark
the X-ray emission linearly spaced from 0.033 to 0.017 ${\rm cts\ arcsec}^{-1}$ (in the
$0.1-2.4\ {\rm keV}$ energy range) and white contours represent $1.5$ to
$4\ \sigma$ (spaced by $0.5\ \sigma$) optical light over-densities from galaxies
above the background level. The light distribution is calculated on a regular
grid (spacing 3\myarcmin 2), where each gridpoint contains the total flux
within 8\myarcmin 0 weighted by a Gaussian with a width of 1\myarcmin 5.  Each
galaxy with $R>17$ (see also Fig.~\ref{fig:rhmagbmr}) is included and the
background level and the $\sigma$ are determined from the gridpoints. The
black square marks a 8\myarcmin 0$\times$8\myarcmin 0 region around the X-ray
peak and is shown enlarged in the {\bf right panel} (8\myarcmin 0 corresponds to
$0.46\ {\rm Mpc}$ at $z=0.05$). The X-ray emission
is reproduced as black contours. The coordinates in the top label 
(given in the
J2000 system) mark the location of the X-ray peak position.  B1 shows a bright
X-ray emission and clear light over-density is seen at the X-ray position. In
the $B-R$ color space we see an over-density of galaxies at around $B-R\approx
2.2$ (see Fig.~\ref{fig:colmag1}) which is very red if it originates from a
galaxy cluster at $z\approx 0.05$.  We consider this case as a good candidate
for a new, previously unidentified galaxy cluster.}
\label{fig:B1}
\end{figure*}

\begin{figure*}
\centering
\resizebox{\hsize}{!}{\includegraphics[angle=-90]{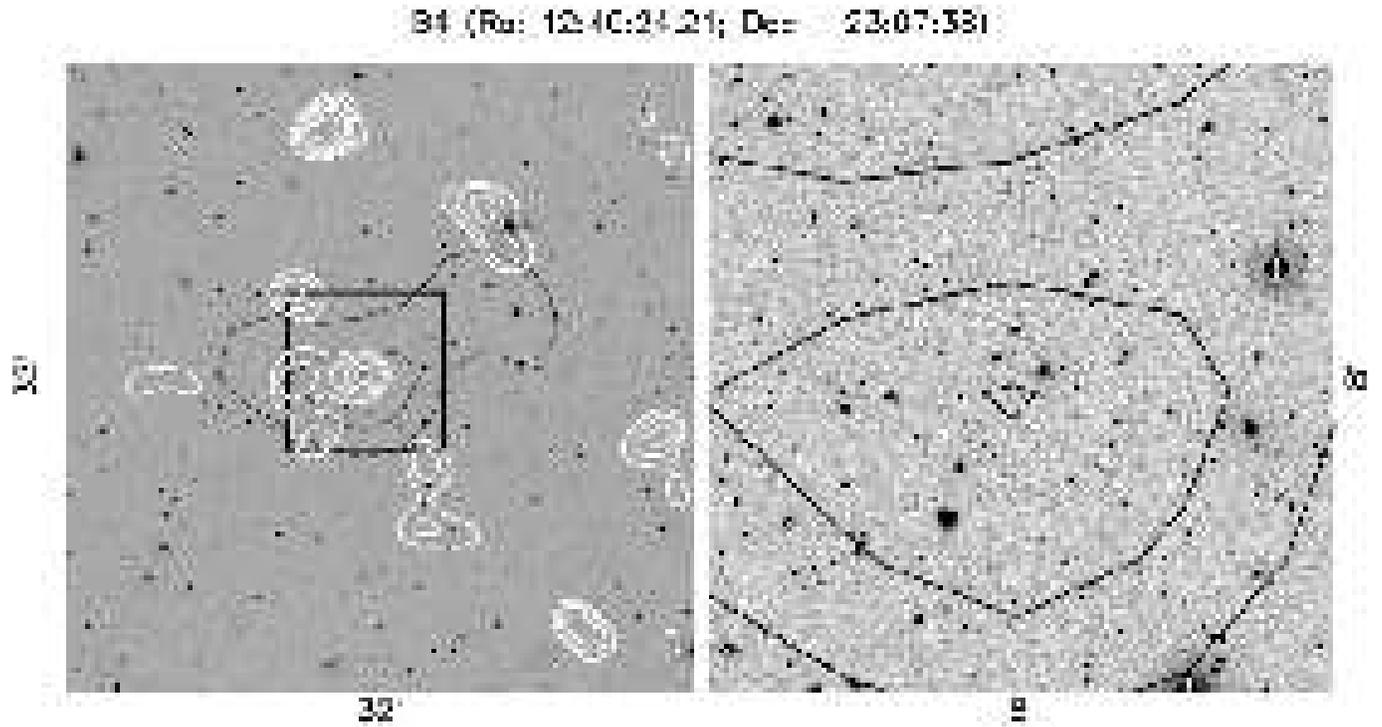}}
\caption{Shown are the X-ray and light distributions around our candidate B4. See Fig.~\ref{fig:B1} for an explanation
of the elements in the figure.
A clear light over-density is seen
at the X-ray peak position. In the $B-R$ color space we see indications of an over-density of galaxies at 
around $B-R\approx 1.6-1.8$ (see Fig.~\ref{fig:colmag1}).
Hence, we consider this case as a good candidate for a new, previously unidentified galaxy cluster.
}
\label{fig:B4}
\end{figure*}

\begin{figure*}
\centering
\resizebox{\hsize}{!}{\includegraphics[angle=-90]{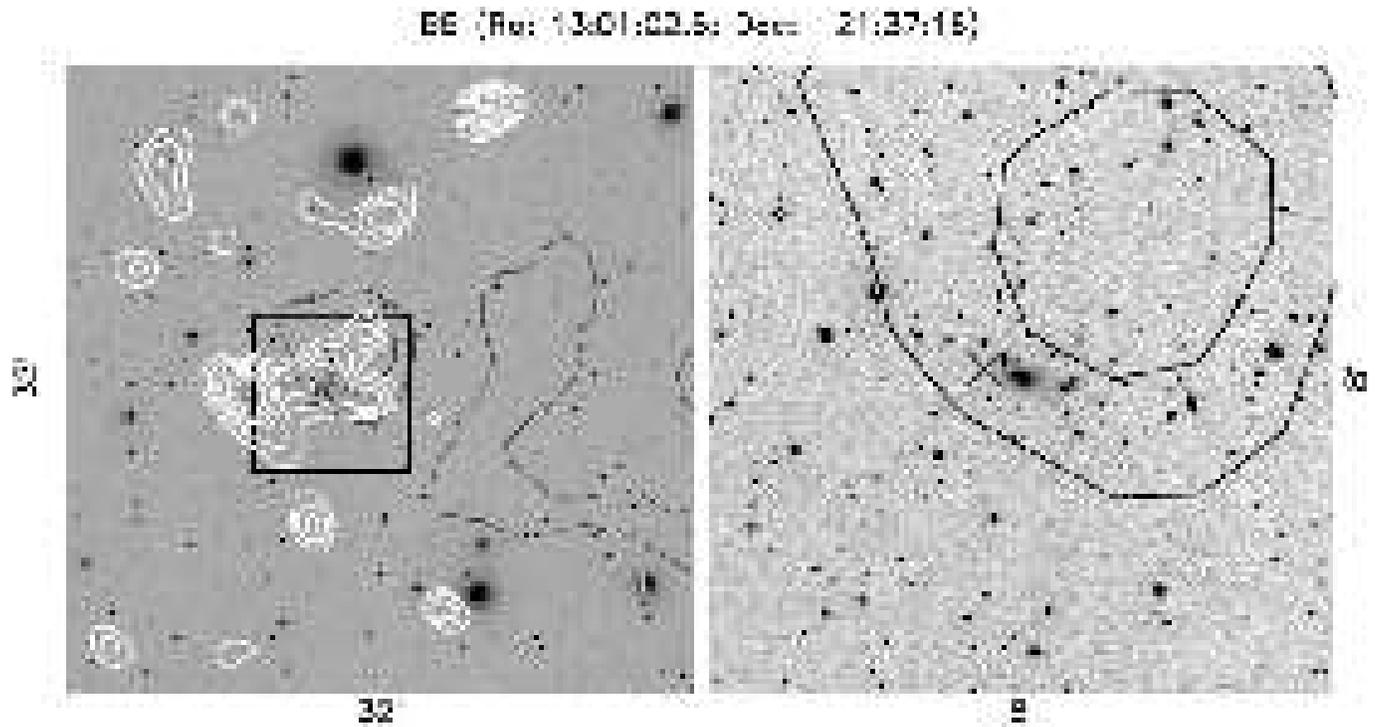}}
\caption{Shown are the X-ray and light distributions around our
candidate B5. See Fig.~\ref{fig:B1} for an explanation of the elements
in the figure. A very strong and elongated light over-density
is seen at the X-ray peak position. In the $B-R$ color space we see a
clear sequence at $B-R\approx 1.5-1.6$ (see
Fig.~\ref{fig:colmag1}). In the course of our work we realized that B5
coincides with the position of the galaxy cluster Abell 3538 (marked
with a cross in the plots). In in this figure, the 8\myarcmin
0$\times$8\myarcmin 0 cutout is centered around the coordinates of the
Abell cluster. There is no measured redshift for it in the literature. }
\label{fig:B5}
\end{figure*}

\begin{figure*}
\centering
\resizebox{\hsize}{!}{\includegraphics[angle=-90]{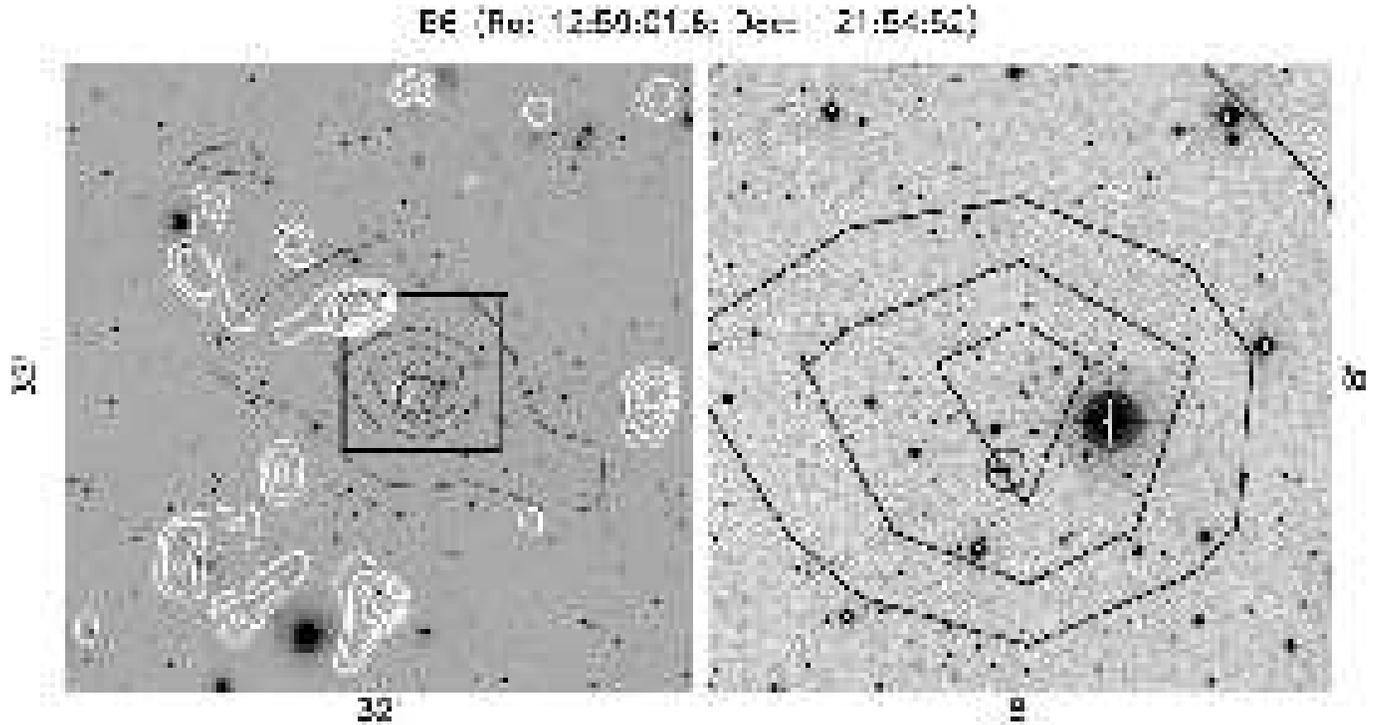}}
\caption{Shown are the X-ray and light distributions around our candidate B6. See Fig.~\ref{fig:B1} for an explanation
of the elements in the figure. 
B6 has a bright X-ray emission and the primary peak seems to
originate from the QSO HE1256$-$2139 ($z=0.146$) at 12:59:02.4; -21:55:38 (J2000) whose position is
marked with a circle. However, the X-ray emission
is elongated in the east-west direction and we identify a strong light over-density around 12:59:29.86;
-21:52:07.4 (J2000) within the X-ray emission area. In the $B-R$ plot we also see indications for a 
sequence at around $B-R\approx 1.5-1.7$ (see Fig.~\ref{fig:colmag1}). We checked that the sequence
does not become more prominent if we include more galaxies to the north-east of the light over-density.
We consider B6 as a good candidate for a new, hitherto unidentified galaxy cluster.}
\label{fig:B6}
\end{figure*}

\begin{figure*}
\centering
\resizebox{\hsize}{!}{\includegraphics[angle=-90]{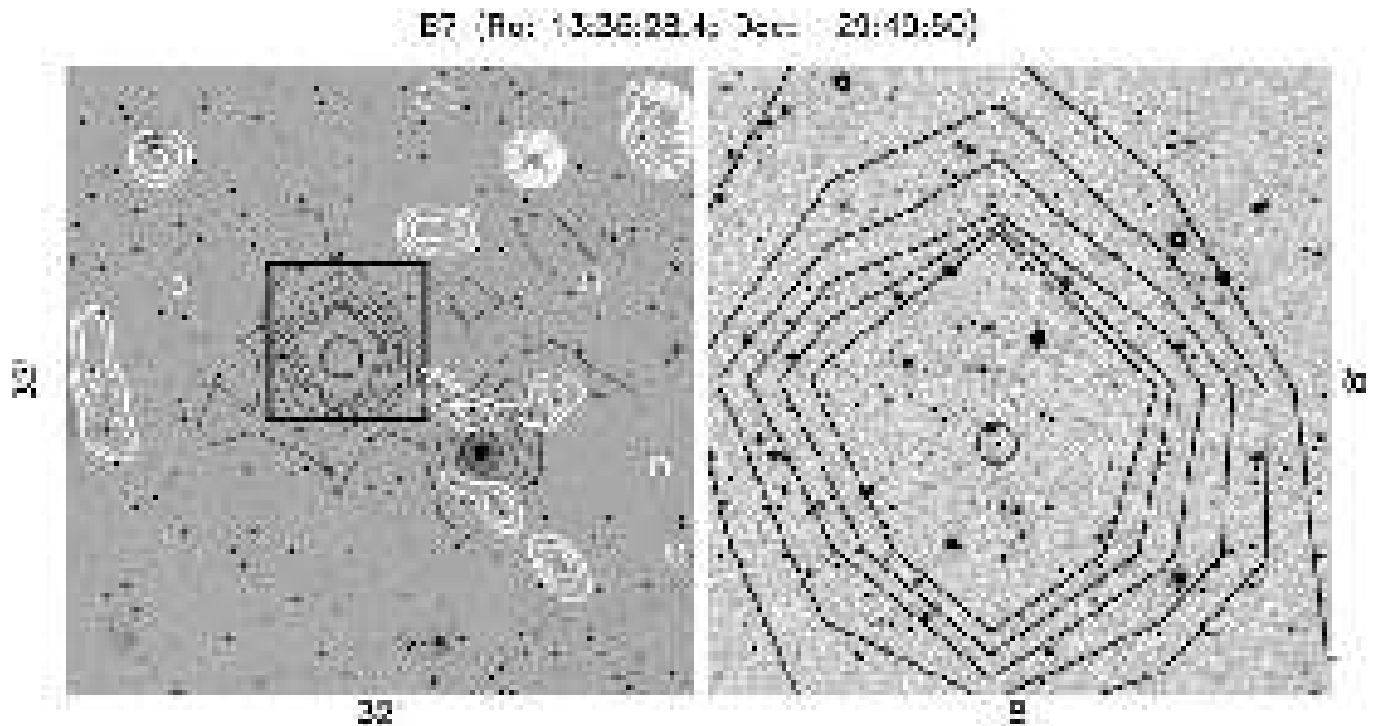}}
\caption{Shown are the X-ray and light distributions around our candidate B7. See Fig.~\ref{fig:B1} for an explanation
of the elements in the figure.
B7 shows a double peak X-ray emission that is probably the superposition of the distributions from the QSO
RBS1291 ($z=0.25$) at 13:35:29.7s;-29:50:39s (J2000) and the bright variable star V347 Hya
at 13:34:57.40; -29:55:24.0 (J2000). Both point sources are marked with a circle. We see only very slight light over-densities 
within the X-ray emission and no signs of a sequence in the $B-R$ diagram (see Fig.~\ref{fig:colmag2}). Hence,
we consider the presence of a new Shapley member in this field as very unlikely.}
\label{fig:B7}
\end{figure*}

\begin{figure*}
\centering
\resizebox{\hsize}{!}{\includegraphics[angle=-90]{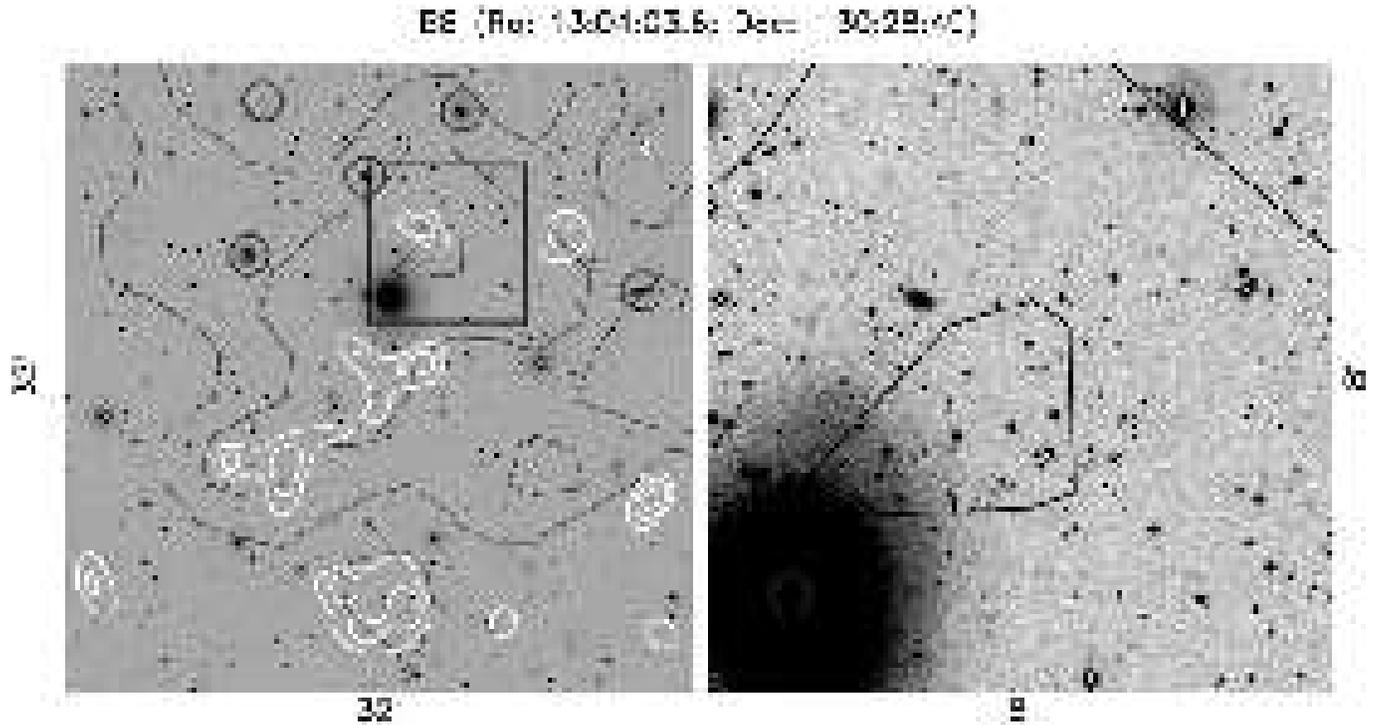}}
\caption{Shown are the X-ray and light distributions around our
candidate B8. See Fig.~\ref{fig:B1} for an explanation of the
elements in the figure.  B8 shows a bright, extremely diffuse 
X-ray emission. We identified several
bright galaxies with measured redshifts between 0.0097 and 0.0116
which are marked with circles. Hence, the X-ray flux most probably
originates from a galaxy group at $z\approx 0.01$. We see a slight
light over-density very close to the X-ray peak but no trends in the
$B-R$ diagram although many bright galaxies populate the region between
$1.5<B-R<2$ (see Fig.~\ref{fig:colmag2}). Hence, we consider this
case as uncertain. We note that the X-ray peak position is listed as
cluster candidate (RXCJ1304.2-3030) within the REFLEX sample; see~\cite{bsg04}.}
\label{fig:B8}
\end{figure*}

\begin{figure*}
\centering
\resizebox{\hsize}{!}{\includegraphics[angle=-90]{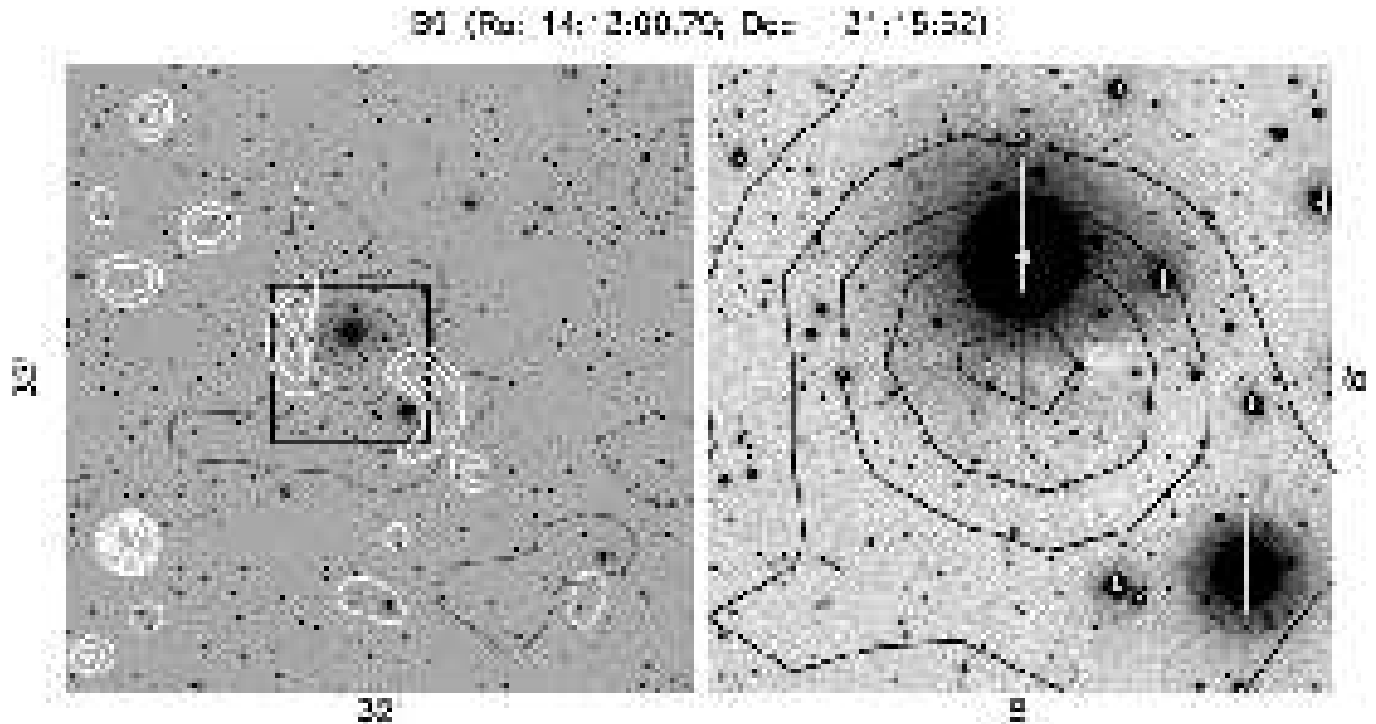}}
\caption{Shown are the X-ray and light distributions around our candidate B9. See Fig.~\ref{fig:B1} for an explanation
of the elements in the figure.
B9 shows a bright X-ray emission. Probably the only source for the observed X-ray emission is the bright 
star close to the X-ray peak. We see slight galaxy over-densities within the emission area and no signs for
galaxy concentrations in the $B-R$ plot (see Fig.~\ref{fig:colmag2}). Hence, we consider it unlikely that
a Shapley member contributes to the X-ray flux in this field.}
\label{fig:B9}
\end{figure*}

\begin{figure*}
\centering
\resizebox{\hsize}{!}{\includegraphics[angle=-90]{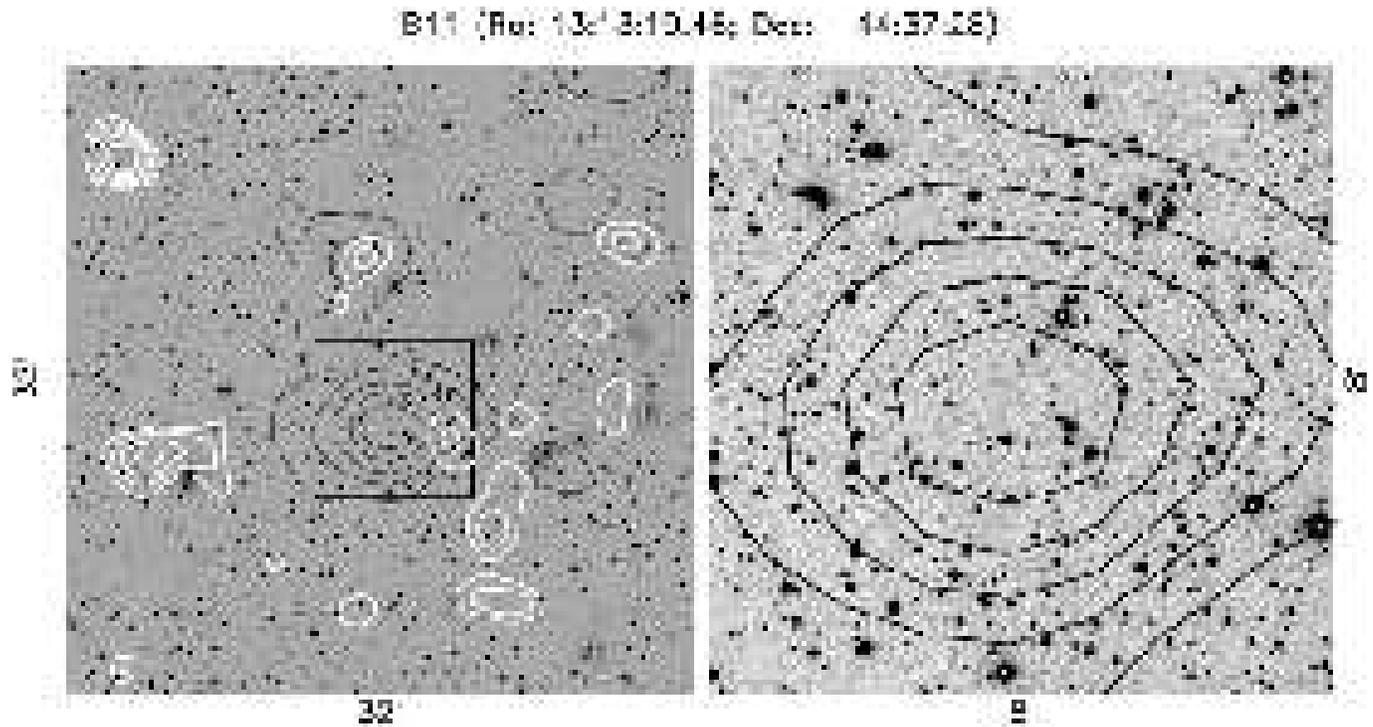}}
\caption{Shown are the X-ray and light distributions around our candidate B11. See Fig.~\ref{fig:B1} for an explanation
of the elements in the figure.
B11 is a field with a high stellar density which significantly hampers
the optical search for galaxy over-densities. We see no indications for the possible presence of a galaxy cluster
in the light or color ($B-R$) distributions (see also Fig.~\ref{fig:colmag2}). 
The most probable explanation for the X-ray emission is the superposition of point sources.}
\label{fig:B11}
\end{figure*}

\begin{figure*}
\centering
\resizebox{\hsize}{!}{\includegraphics[angle=-90]{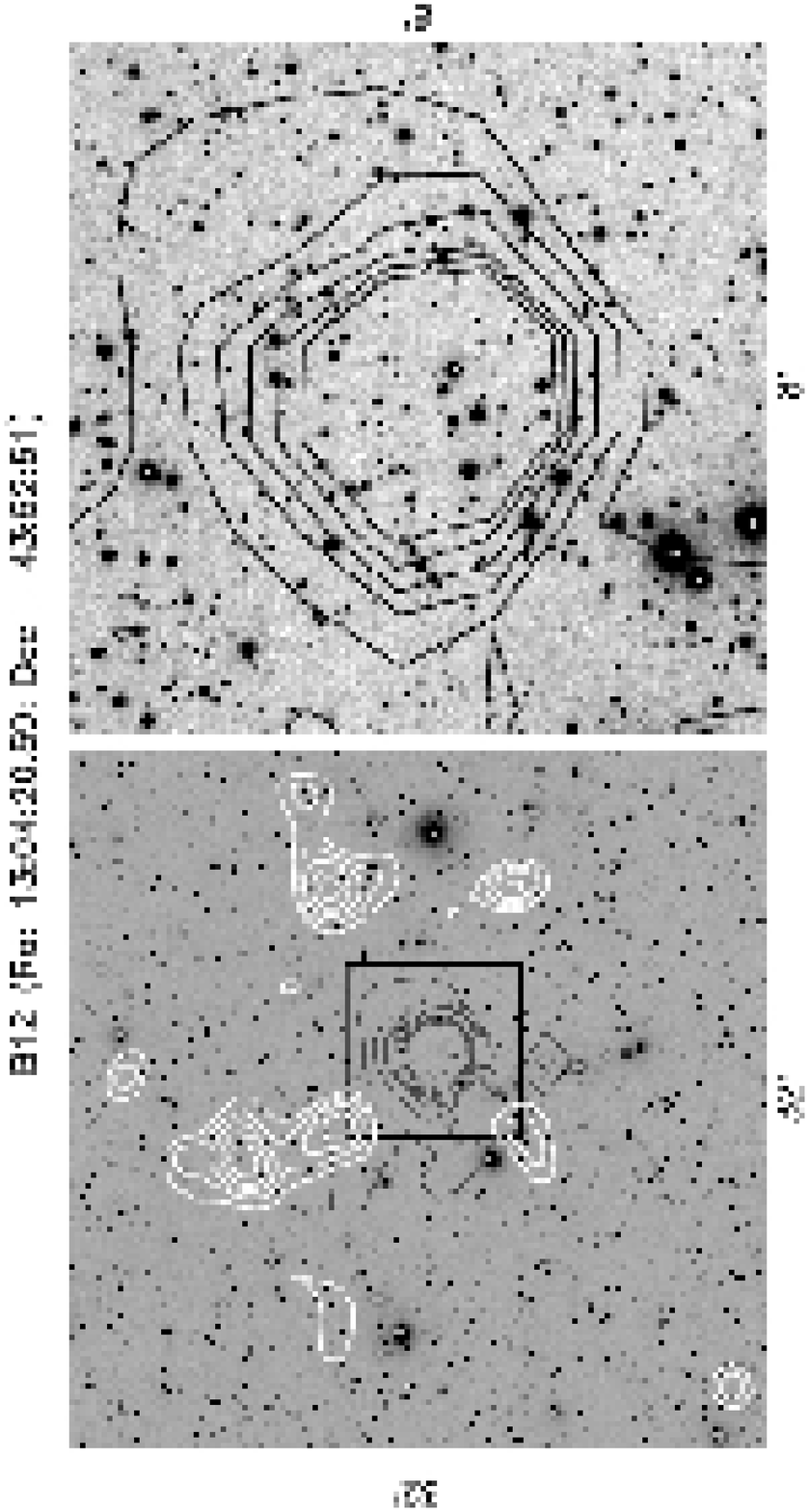}}
\caption{Shown are the X-ray and light distributions around our candidate B12. See Fig.~\ref{fig:B1} for an explanation
of the elements in the figure.
B12 shows a bright an extended
X-ray emission. The case of B12 also lies within a stellar field as B11 and our conclusions
are the same as for that field.}
\label{fig:B12}
\end{figure*}

\begin{figure*}
\centering
\resizebox{\hsize}{!}{\includegraphics[angle=-90]{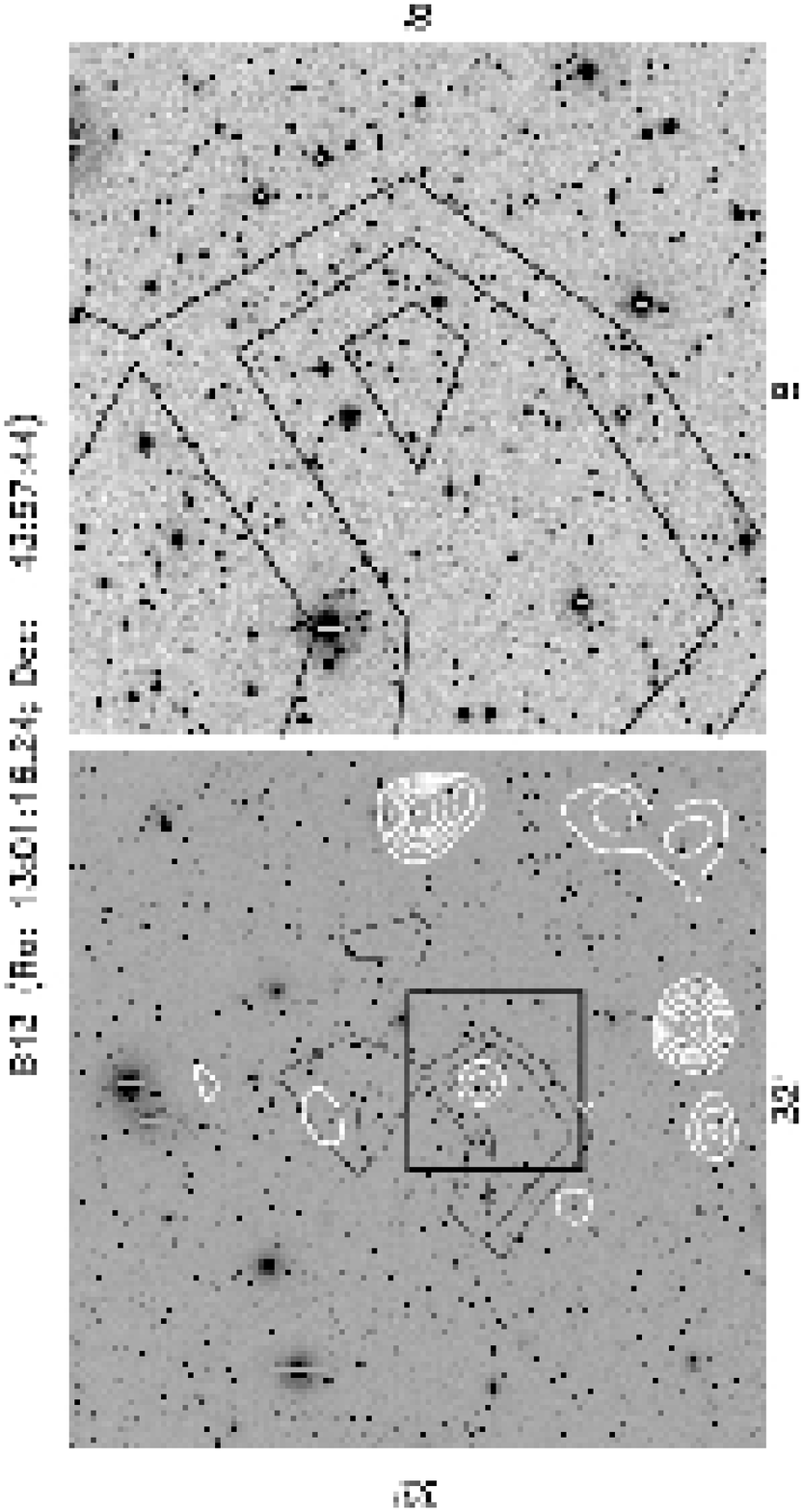}}
\caption{Shown are the X-ray and light distributions around our candidate B13. See Fig.~\ref{fig:B1} for an explanation
of the elements in the figure.
B8 shows an extended
X-ray emission. The case of B13 lies within a stellar field as B11 and our conclusions
are the same as for that field.}
\label{fig:B13}
\end{figure*}

\begin{figure*}
\centering
\resizebox{\hsize}{!}{\includegraphics[angle=-90]{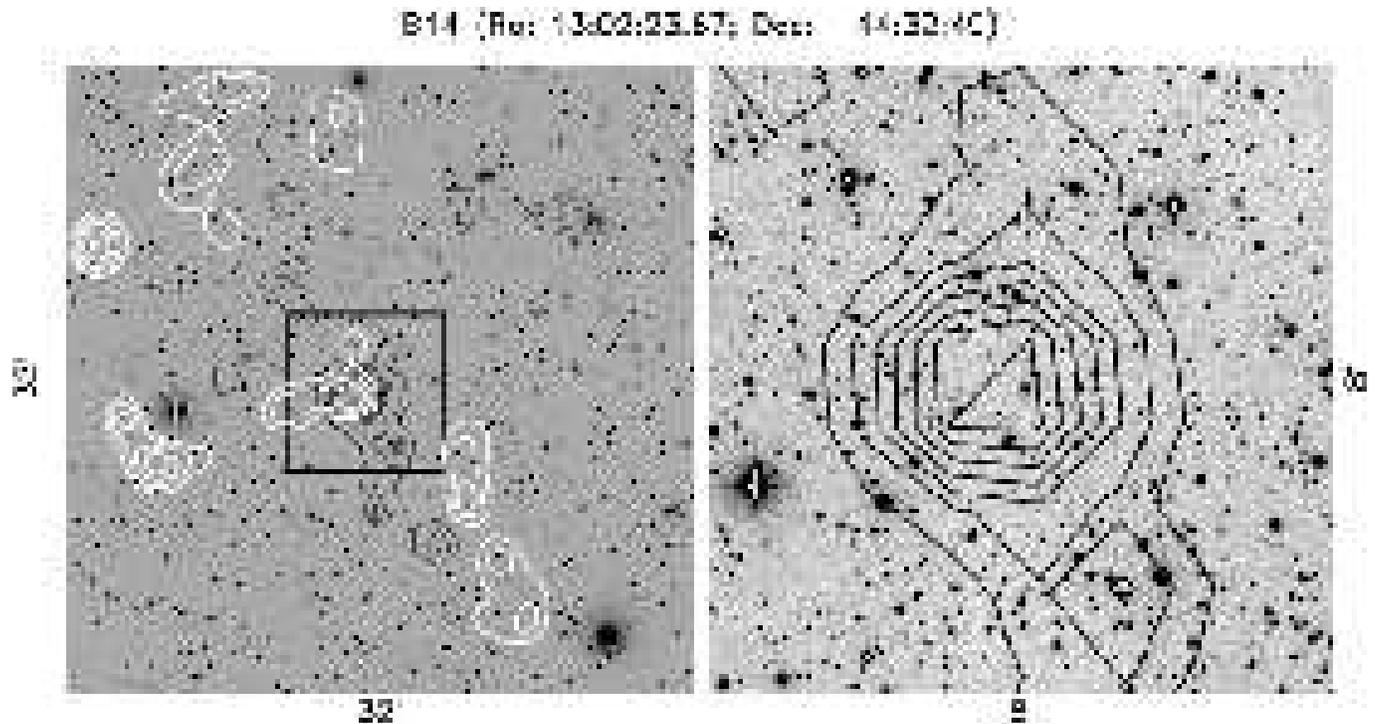}}
\caption{Shown are the X-ray and light distributions around our candidate B14. See Fig.~\ref{fig:B1} for an explanation
of the elements in the figure.
B14 shows an elongated X-ray emission. 
Similar to B11, B12 and B13 the location within a stellar field makes a quantitative analysis in the optical
difficult. Because of the good X-ray detection and an extended over-density directly at the X-ray peak we classify
this case as uncertain. No indications for a cluster sequence are seen  in the $B-R$ diagram; see 
Fig.~\ref{fig:colmag3}.}
\label{fig:B14}
\end{figure*}

\begin{figure*}
\centering
\resizebox{\hsize}{!}{\includegraphics{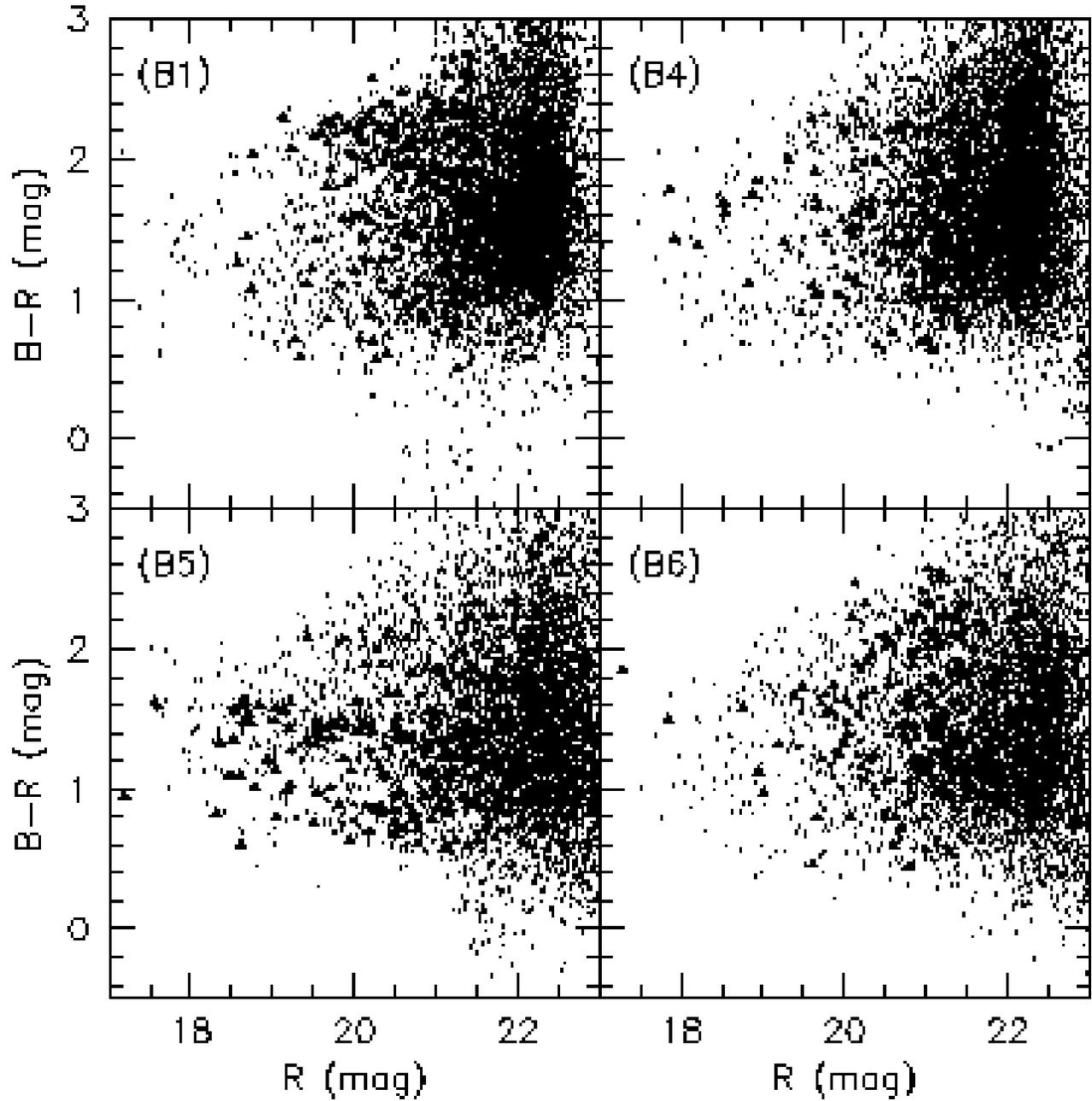}}
\caption{Shown are color-magnitude diagrams for our candidates B1, B4, B5 and B6. Dots represent all
galaxies within the corresponding WFI fields. Filled triangles represent the galaxies in a 
8\myarcmin 0$\times$ 8\myarcmin 0 around the X-ray peak positions (see Figs.~\ref{fig:B1}-\ref{fig:B6}).
For rich galaxy clusters at $z\approx 0.05$ we would expect a Red-Cluster Sequence from elliptical
galaxies around $1.5<B-R<2$ (see also Fig.~\ref{fig:rhmagbmr}).}
\label{fig:colmag1}
\end{figure*}

\begin{figure*}
\centering
\resizebox{\hsize}{!}{\includegraphics{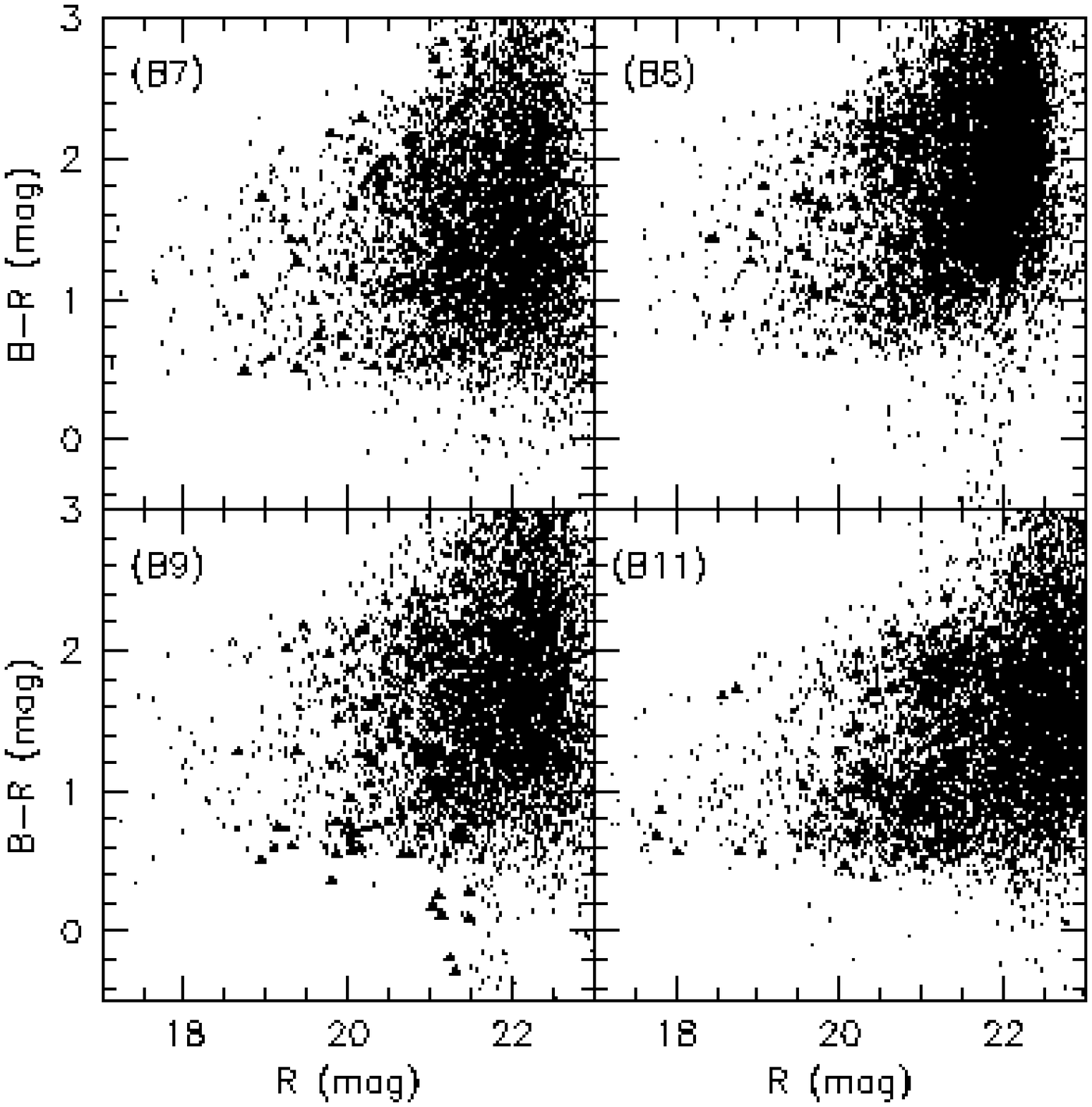}}
\caption{Shown are color-magnitude diagrams for our candidates B7, B8, B9 and B11. For an explanation of the plot see the caption from Fig.~\ref{fig:colmag1}.}
\label{fig:colmag2}
\end{figure*}

\begin{figure*}
\centering
\resizebox{\hsize}{!}{\includegraphics{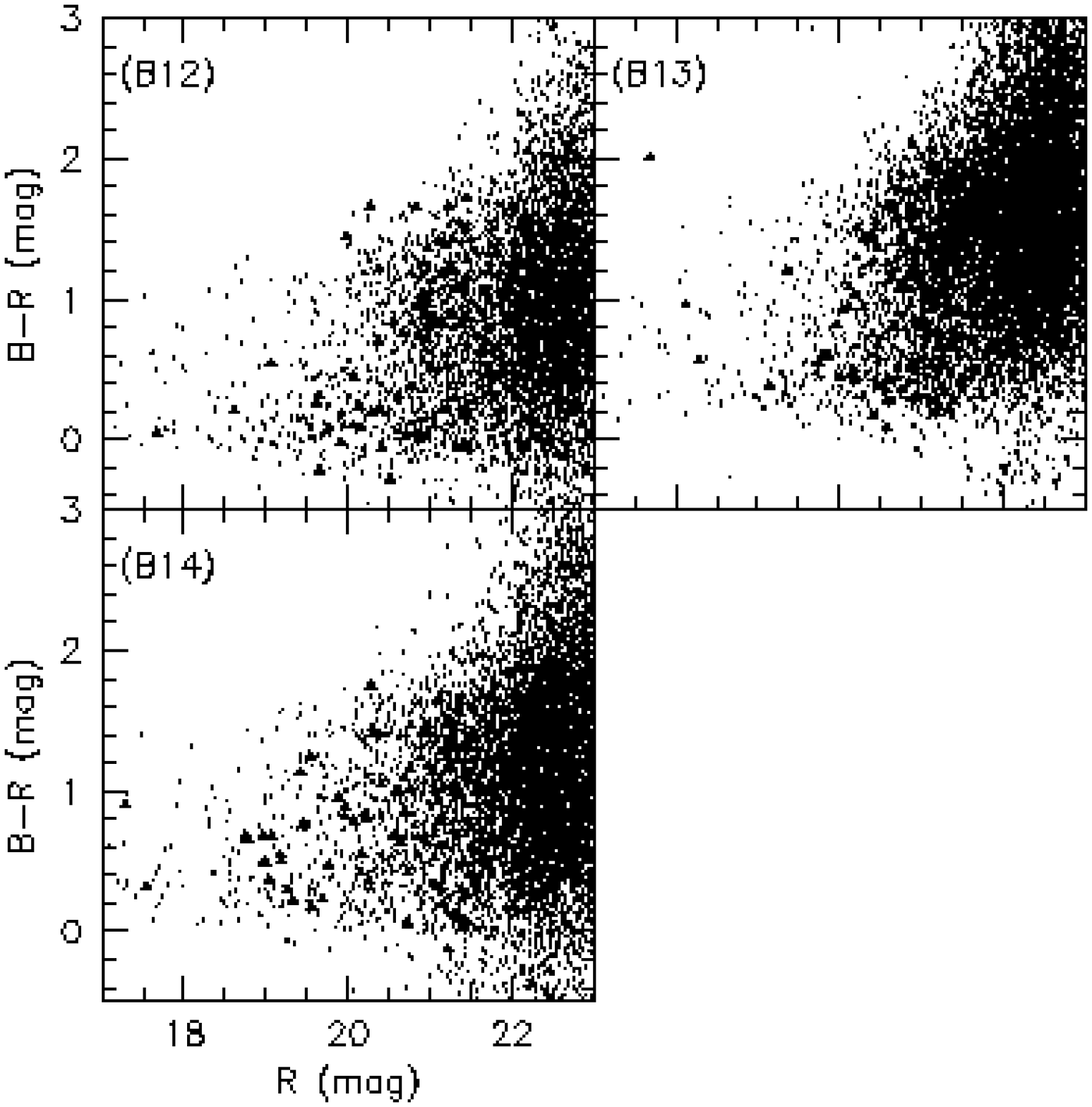}}
\caption{Shown are color-magnitude diagrams for our candidates 
B7, B8, B9 and B11. For an explanation of the plot see the
 caption from Fig.~\ref{fig:colmag1}.}
\label{fig:colmag3}
\end{figure*}

In this Appendix we discuss the analysis of the optical data for our
new Shapley SC candidates B1, B4-B9 and B11-B14 (see
Sect.~\ref{sec:new_det} and Table~\ref{tab:ris_fin}). We extract
object catalogs from our $B$ and $R$ WFI images by using SExtractor~\citep{Ber96} 
in the double image mode, i.e. we use the deeper
$R$-band image to detect objects (we considered all sources having at
least 4 pixels with $2\sigma$ over the sky background noise) and
determined aperture magnitudes in $R$ and $B$ (we used a fixed
aperture with a diameter of 6\myarcsec 0) around these positions.
Fig.~\ref{fig:rhmagbmr} explains how we separate stars and galaxies
and in which area of our color-magnitude diagrams we search for
a Red Cluster Sequence of potential Shapley SC members.
With our selection
criteria, the galaxy number counts start to drop at $R\approx 23$
which should be sufficient for the detection of massive galaxy
clusters at $z\approx 0.05$.
The results of our analysis of all fields is shown in 
Figs.~\ref{fig:B1}-\ref{fig:colmag3}.
\clearpage

\end{document}